\newcommand*{\ATLASLATEXPATH}{}

\pdfoutput=1
\documentclass[cernpreprint, texlive=2016, UKenglish]{\ATLASLATEXPATH atlasdoc}

\usepackage{\ATLASLATEXPATH atlaspackage}
\usepackage{\ATLASLATEXPATH atlasbiblatex}
\usepackage{\ATLASLATEXPATH atlascontribute}
\usepackage{\ATLASLATEXPATH atlasphysics}

\newcommand{\xb}{\mbox{$x_{\mathrm{A}}$}}
\newcommand{\Et}{\mbox{$E_{\mathrm{T}}$}}
\newcommand{\conetwo}{\mbox{$C_{12}$}}
\newcommand{\wonetwo}{\mbox{$W_{12}$}}
\newcommand{\cppb}{\mbox{$\rho_{W}^{\mathrm{pPb}}$}}
\newcommand{\ionetwo}{\mbox{$I_{12}$}}
\newcommand{\ippb}{\mbox{$\rho_{I}^{\mathrm{pPb}}$}}
\newcommand{\sqrtsnn}{\mbox{$\sqrt{s_{_\text{NN}}}$}}
\newcommand{\sqrts}{\mbox{$\sqrt{s}$}}
\newcommand{\pythia}{\textsc{Pythia}}

\newcommand{\pythiaeight}{\textsc{Pythia}8}
\newcommand{\herwig}{\textsc{Herwig}++}
\newcommand{\RFour}{\mbox{$R = 0.4$}}
\newcommand{\Dphi}{\mbox{$\Delta \phi$}}
\newcommand{\Deta}{\mbox{$\Delta \eta$}}

\newcommand{\pPb}{\mbox{$p$+Pb}}
\newcommand{\ptone}{\mbox{$p_{\mathrm{T,1}}$}}
\newcommand{\pttwo}{\mbox{$p_{\mathrm{T,2}}$}}

\newcommand{\pttruth}{\mbox{$p_{\mathrm{T}}^{\mathrm{truth}}$}}
\newcommand{\ptreco}{\mbox{$p_{\mathrm{T}}^{\mathrm{reco}}$}}
\newcommand{\etatruth}{\mbox{$\eta_{\mathrm{truth}}$}}
\newcommand{\etareco}{\mbox{$\eta_{\mathrm{reco}}$}}
\newcommand{\phitruth}{\mbox{$\phi_{\mathrm{truth}}$}}
\newcommand{\phireco}{\mbox{$\phi_{\mathrm{reco}}$}}

\newcommand{\ystar}{\mbox{$y^{*}$}}

\newcommand{\ystarone}{\mbox{$y^{*}_{\mathrm{1}}$}}
\newcommand{\ystartwo}{\mbox{$y^{*}_{\mathrm{2}}$}}

\AtlasTitle{ Dijet azimuthal correlations and conditional yields in  $p\!p$ and $p$+Pb collisions at $\sqrt{s_{_\text{NN}}}$~=~5.02~\TeV\ with the ATLAS detector}

\AtlasAbstract{
	This paper presents a measurement of forward--forward and forward--central dijet azimuthal angular correlations and conditional yields in proton--proton ($pp$) and proton--lead ($p$+Pb) collisions as a probe of the nuclear gluon density in regions where the fraction of the average momentum per nucleon carried by the parton entering the hard scattering is low. In these regions, gluon saturation can modify the rapidly increasing parton distribution function of the gluon. The analysis utilizes 25~pb$^{-1}$ of $pp$ data and 360~$\mu \mathrm{b}^{-1}$ of $p$+Pb data, both at $\sqrt{s_{_\text{NN}}}$~=~5.02~\TeV, collected in 2015 and 2016, respectively, with the ATLAS detector at the LHC. The measurement is performed in the center-of-mass frame of the nucleon--nucleon system in the rapidity range between $-4.0$ and 4.0 using the two highest transverse momentum jets in each event, with the highest transverse momentum jet restricted to the forward rapidity range. No significant broadening of azimuthal angular correlations is observed for forward--forward or forward--central dijets in $p$+Pb compared to $pp$ collisions. For forward--forward jet pairs in the proton-going direction, the ratio of conditional yields in $p$+Pb collisions to those in $pp$ collisions is suppressed by approximately 20\%, with no significant dependence on the transverse momentum of the dijet system. No modification of conditional yields is observed for forward--central dijets. 
}

\AtlasRefCode{HION-2017-06}
\AtlasJournal{PRC}
\PreprintIdNumber{CERN-EP-2018-324}
\AtlasJournalRef{Phys. Rev. C 100 (2019) 034903}
\AtlasDOI{10.1103/PhysRevC.100.034903}

\hypersetup{pdftitle={ATLAS document},pdfauthor={The ATLAS Collaboration}}

\begin{document}

\maketitle

\section{Introduction}
\label{sec:intro}
Studies of particle collisions at accelerators have contributed significantly to an improved understanding of the strong interaction in quantum chromodynamics (QCD) and to the knowledge of the parton distribution functions (PDFs) of the proton. Global QCD analyses of structure functions in deep-inelastic lepton--nucleon scattering at HERA, as well as jet and hadron cross-sections at the LHC, Tevatron, and RHIC were performed in a wide kinematic range, providing several new sets of PDFs with the highest degree of precision reached so far~\cite{Dulat:2015mca,Ball:2017nwa,Harland-Lang:2014zoa,Abramowicz:2015mha}. These analyses  constrain quark and gluon contributions over a wide range of the Bjorken variable $x$: the longitudinal momentum fraction of a nucleon carried by its constituent partons. From these measurements, the gluon distribution in the proton is found to rise rapidly for decreasing $x$. Unitarity requires that the first moment of the gluon momentum distribution remains finite. Therefore, the steep rise at low $x$ must change at some $x$ value; this phenomenon is known as \textit{saturation}~\cite{Albacete:2014fwa}. 

The search for the onset of saturation was a major scientific goal with deuteron--gold and gold--gold collisions at RHIC~\cite{Adler:2003au, Wang:2007zv, Adare:2011sc}, where the sensitivity to saturation effects was increased due to the enhancement of the nuclear gluon density in the Lorentz-contracted  nucleus~\cite{Blaizot:2016qgz}. These measurements were able to probe the parton longitudinal-momentum fraction of the nucleon in the nucleus down to  $\xb\sim10^{-3}$. Currently, the gluon nuclear PDFs have large uncertainties at low \xb~\cite{Kovarik:2015cma, Eskola:2016oht}, and additional data in this region would help to further constrain them. A mid-rapidity measurement of jet-production rates at RHIC found no significant modification in deuteron--gold collisions compared to proton--proton (\pp) collisions~\cite{Adare:2015gla}. Recent analyses at the LHC have been performed in the proton-going direction of proton--lead (\pPb) collisions and at higher center-of-mass energies, allowing a lower value of \xb\ to be probed for the lead nucleus. The ALICE measurements of cross-sections for charged-jet production and dijet azimuthal angular correlations at mid-rapidity did not find significant modifications in \pPb\ collisions compared to \pp\ collisions~\cite{Adam:2015xea, Adam:2015hoa}. The ATLAS and CMS analyses of inclusive jet production also did not find significant evidence of nuclear modification~\cite{ATLAS:2014cpa, Khachatryan:2016xdg}. Another approach to probe gluon saturation in nuclear gluon densities was proposed in the framework of the Color Glass Condensate (CGC) model~\cite{Kharzeev:2004bw} by studying the modifications of dijet azimuthal angular distributions in \pp\ and \pPb\ collisions at forward rapidities at \xb\ down to $10^{-5}$~\cite{vanHameren:2014lna}. For back-to-back dijets, the gluon field in the lead nucleus is probed at low momentum where saturation effects are expected to be large~\cite{Kutak:2013yga, vanHameren:2019ysa}.

In this paper, a measurement of  azimuthal correlations between leading and subleading jets in \pp\ and \pPb\ collisions at $\sqrt{s_{_\text{NN}}}$~=~5.02~\TeV\ is presented. The measurement is performed in intervals of the jet center-of-mass rapidity\footnote{ ATLAS uses a right-handed coordinate system with its origin at the nominal interaction point (IP) in the center of the detector and the $z$-axis along the beam pipe. The $x$-axis points from the IP to the center of the LHC ring, and the $y$-axis points upward. Cylindrical coordinates $(r,\phi)$ are used in the transverse plane, $\phi$ being the azimuthal angle around the beam pipe. For the \pPb\ collisions, the incident Pb beam traveled in the $+z$ direction. The pseudorapidity is defined in terms of the polar angle $\theta$ as $\eta=-\ln\tan(\theta/2)$. Angular distance is measured in units of $\Delta R \equiv \sqrt{(\Delta\eta)^{2} + (\Delta\phi)^{2}}$ with $\Delta\eta$ and $\Delta\phi$ defined as the differences between two directions in pseudorapidity and azimuth. Rapidity is defined in terms of energy and momentum of a particle or jet as $y=(1/2)\ln[(E+p_{z})/(E-p_{z})]$.} 
$\ystar = y - \Delta y$, where $y$ is the jet rapidity in the laboratory frame, and $\Delta y$ is the rapidity shift of the center-of-mass frame relative to the laboratory frame. This shift results from the different energy of the proton-beam with respect to the Pb beam in \pPb\ collisions. The leading jet  has the highest transverse momentum (\ptone) in the event and is required to be in the forward proton-going direction; otherwise, the event is not considered. The subleading jet has the second-highest transverse momentum (\pttwo) in the event and its rapidity range is not restricted. The center-of-mass rapidities of the leading and subleading jets are \ystarone\ and \ystartwo, respectively. This measurement of dijets can probe the \xb\ range between $10^{-4}$ and $10^{-3}$ in the lead nucleus. The azimuthal angular correlation distributions \conetwo, which are normalized to the number of forward $(2.7<\ystarone<4.0)$ leading jets  $N_{1}$ in a given \ptone\ interval, are defined as:
\begin{equation}
\conetwo(\ptone, \pttwo, \ystarone, \ystartwo)=\frac{1}{N_{1}}\frac{dN_{12}}{d\Dphi},\nonumber
\end{equation}
where $N_{12}$ is the number of dijets, and \Dphi\ is the azimuthal angle between the leading and subleading jets. The \conetwo\ distributions are fitted and their widths \wonetwo\ defined by the root-mean-square of the fit function: $\wonetwo(\ptone, \pttwo, \ystarone, \ystartwo) = \mathrm{RMS}(\conetwo)$.

In addition to dijet azimuthal angular distributions, the dijet conditional yields \ionetwo\ are measured and defined as:
\begin{equation}
\ionetwo(\ptone, \pttwo, \ystarone, \ystartwo)=\frac{1}{N_{1}}  \frac{d^{4}N_{12}}{d\ystarone d\ystartwo d\ptone d\pttwo}  .\nonumber
\end{equation} 

The azimuthal angular correlations and conditional yields evaluated in \pPb\ and \pp\ collisions are compared and the ratios in \wonetwo\ and \ionetwo\ between the two systems are calculated as:
\begin{equation}
\cppb(\ptone, \pttwo, \ystarone, \ystartwo)=\frac{W_{12}^{\mathrm{pPb}}}{W_{12}^{\mathrm{pp}}} \ \ , \ \ \ippb(\ptone, \pttwo, \ystarone, \ystartwo)=\frac{I_{12}^{\mathrm{pPb}}}{I_{12}^{\mathrm{pp}}}.\nonumber
\end{equation}
To define a phase space that better suits next-to-leading-order calculations, a minimum $\Delta \pT=\ptone - \pttwo$ is required for the dijets~\cite{Potter:1999gg,Klasen:1995xe,Frixione:1997ks}. However, techniques such as Sudakov resummation~\cite{Bury:2017jxo} can take into account the absence of $\Delta \pT$ requirements. Also, comparisons with fixed-order calculations and soft-gluon resummation, which involve transverse-momentum-dependent PDFs, instead of collinear PDFs, are better suited to scenarios not placing any minimum $\Delta \pT$ requirement on the dijets. The results of the measurement are therefore presented both without any requirement on $\Delta\pT$ and with a requirement of $\Delta\pT>3$~\GeV.

\section{Experimental setup}
\label{sec:setup}
The measurements presented here are performed using the ATLAS calorimeter, trigger, and data acquisition systems \cite{Aad:2008zzm}. The calorimeter system consists of a sampling lead/liquid argon (LAr) electromagnetic calorimeter covering $|\eta|<3.2$, a steel/scintillator sampling hadronic calorimeter covering $|\eta|<1.7$, a LAr hadronic calorimeter covering $1.5<|\eta|<3.2$, and two LAr forward calorimeters (FCal) covering $3.2<|\eta|<4.9$. The electromagnetic calorimeters are segmented longitudinally in shower depth into three layers plus an additional presampler layer and have a granularity that varies with the layer and pseudorapidity, and which is also much finer than that of the hadronic calorimeter.  The hadronic calorimeter has three longitudinal sampling layers and comprises the Tile barrel and extended barrel hadronic calorimeters covering $|\eta|<1.7$, and the hadronic endcap calorimeter (HEC) covering $1.5<|\eta|<3.2$. The minimum-bias trigger scintillators detect particles over $2.1<|\eta|<3.9$ using two azimuthally segmented counters placed at $z=\pm 3.6$~m. There are 12 measurements per counter. Each counter provides measurements of both the  pulse heights and the arrival times of energy deposits from each segment.

A two-level trigger system was used to select the \pp\ and \pPb\ collisions. The first level is the level-1 (L1) hardware-based trigger implemented with custom electronics. The second level is the software-based high-level trigger (HLT). Jet events were selected by the HLT with input from the L1 jet and transverse-energy triggers in \pp\ collisions, and minimum-bias trigger in \pPb\ collisions. The two L1 transverse-energy triggers used in \pp\ collisions required the total transverse energy measured in the calorimeters to be greater than 5~\GeV\ and 10~\GeV, respectively. The L1 jet trigger used in \pp\ collisions required a jet to exceed transverse-energy thresholds ranging from 12~\GeV\ to 20~\GeV. The L1 minimum-bias trigger selected \pPb\ events with at least one hit in the minimum-bias trigger scintillator counters on each side of the IP. The HLT jet trigger employed a jet reconstruction algorithm similar to that applied in the offline analysis and selected events containing jets that exceeded a transverse-energy threshold of 15~\GeV\ in \pPb\ collisions and thresholds ranging from 25 to 85~\GeV\ in \pp\ collisions. In both the \pp\ and \pPb\ collisions, the highest-threshold jet trigger sampled the full delivered luminosity, and jet triggers with lower thresholds were prescaled\footnote{The prescale indicates which fraction of events that passed the trigger selection was selected for recording by the data acquisition system.} and sampled a fraction of delivered luminosity. Both the forward ($3.2<|\eta|<4.9$) and central  ($|\eta|<3.2$) jet triggers are used in this measurement. 

\section{Data sets and event selection}
\label{sec:data}
A total of 25~pb$^{-1}$ of \sqrts~=~5.02~\TeV\ \pp\ data from 2015 with two equal-energy proton beams is used. During \pp\ data taking, the average number of interactions per bunch crossing  varied from 0.6 to 1.3. 

The \pPb\ data used in this analysis were recorded in 2016 with the LHC configured with a 4 TeV proton-beam and a 1.57~\TeV\ per nucleon Pb beam, producing collisions with \sqrtsnn~=~5.02~\TeV\ and $\Delta y=0.465$. 
The polar angle $\theta$ was $\pi$ for the proton-beam, and zero for the Pb beam. However, in order to be consistent with previous measurements~\cite{ATLAS:2014cpa,Aaboud:2017tke}, the proton-going direction is defined to have positive rapidity in this measurement. The total \pPb\ integrated luminosity is 360~$\mu$b$^{-1}$. During the \pPb\ data taking the average number of \pPb\ interactions per bunch crossing was 0.03. In \pPb\ and \pp\ collisions, events are required to have a reconstructed vertex. Only events taken during stable beam conditions and satisfying detector and data-quality requirements are considered.

The performance of ATLAS in measuring azimuthal angular correlations and conditional yields in both the  \pp\ and \pPb\ data samples was evaluated with a 5.02~\TeV\ \pp\ Monte Carlo (MC) sample simulated using \pythia~8.212~\cite{Sjostrand:2014zea}. Hard-scattering \pp\ events generated with the A14~\cite{ATLAS2014021} set of tuned parameters and the NNPDF23LO PDF set~\cite{Ball:2012cx} were used. The detector response was simulated using GEANT4~\cite{Agostinelli:2002hh,Aad:2010ah}. The \pp\ MC samples used for this analysis contain approximately 12 million events. Corresponding \pPb\ MC samples were obtained by overlaying signal from \pp\ MC simulation with minimum-bias data events from \pPb\ collisions. These simulated 5.02~\TeV\ \pp\ events used in the overlay procedure were generated with the same set of tuned parameters as for the \pp\ MC sample but with a rapidity shift equivalent to that in the \pPb\ collisions. The simulated hits are combined with those from the data event and used as input to the jet reconstruction. Additionally, a \herwig~\cite{Bahr:2008pv} MC simulation of approximately 5.6 million 5.02~\TeV\ \pp\ events was used for performance studies. The \pPb\ MC samples are weighted at the event level to reproduce the FCal \Et\ distribution in the \pPb\ data. 

\section{Jet selection and reconstruction}
\label{sec:reconstruction}
Jets in \pp\ and \pPb\ collisions are reconstructed using the techniques described in Ref~\cite{ATLAS:2014cpa,Aaboud:2018twu}, which are briefly summarized here. The jet reconstruction is first run in the four-momentum recombination mode, on $\Delta \eta \times \Delta \phi = 0.1\times 0.1$ calorimeter towers with the \antikt\ algorithm~\cite{Cacciari:2008qp} with radius parameter $R=0.4$. Energies in the towers are obtained by summing the energies of calorimeter cells at the electromagnetic energy scale within the tower boundaries. Then, an iterative procedure is used to estimate the layer- and $\eta$-dependent underlying event (UE) transverse-energy density, while excluding the regions populated by jets. The UE transverse energy is subtracted from each calorimeter tower and the four-momentum of the jet is updated accordingly. Then, a jet $\eta$~- and \pT-dependent  correction factor derived from the simulation samples is applied to correct for the calorimeter response. An additional correction based on in situ studies of the transverse-momentum balance of jets recoiling against photons, $Z$ bosons, and jets in other regions of the calorimeter is applied ~\cite{Aad:2011he,Aaboud:2017jcu}.

Jets are selected in the transverse-momentum range $28<\pT<90$~\GeV\ and the center-of-mass rapidity range $|\ystar|<4.0$. These selections guarantee the largest symmetric overlap between the two colliding systems for which most forward jets can be reconstructed using the FCal with full coverage for \RFour\ jets. All reconstructed jets are required to  have a $\pt > 28$ GeV such that the jet trigger efficiency is greater than 99\%. As a result, no trigger efficiency correction is applied. During the \pPb\ data taking, part of the HEC was disabled in the pseudorapidity and azimuthal intervals $1.3<\eta<3.2$ and $-\pi<\phi<-\pi/2$. Reconstructed dijets where the subleading jet area overlaps with the disabled HEC region are excluded from the analysis in \pPb\ data and MC samples.  

The MC samples are used to evaluate the jet reconstruction performance and to correct the measured distributions for detector effects. This is done independently for \pp\ and \pPb\ collisions. In the MC samples, the generator-level jets are reconstructed from stable particles\footnote{Stable particles are defined as particles with a mean lifetime $\tau>0.3\times 10^{-10}$~s.} excluding muons and neutrinos, with the \antikt\ algorithm with radius parameter \RFour. Using the pseudorapidity and azimuthal angles \etatruth, \phitruth, \etareco, and \phireco\ of the generated and reconstructed jets, respectively, generator-level jets are matched to reconstructed jets by requiring $\Delta R < 0.2$.

The efficiency for reconstructing jets in \pp\ and \pPb\ collisions is evaluated using the \pythiaeight\ MC samples by determining the probability of finding a reconstructed jet associated with a generator-level jet. The jet reconstruction efficiency is greater than 99\% for jets with $\pT>30$~\GeV\ and decreases to 95\% at a jet $\pt = 28$~\GeV. The jet reconstruction efficiency exhibits a small variation with rapidity.

\begin{figure}
	\centerline{
		\begin{tabular}{ccc}
			\includegraphics[width=0.45\textwidth]{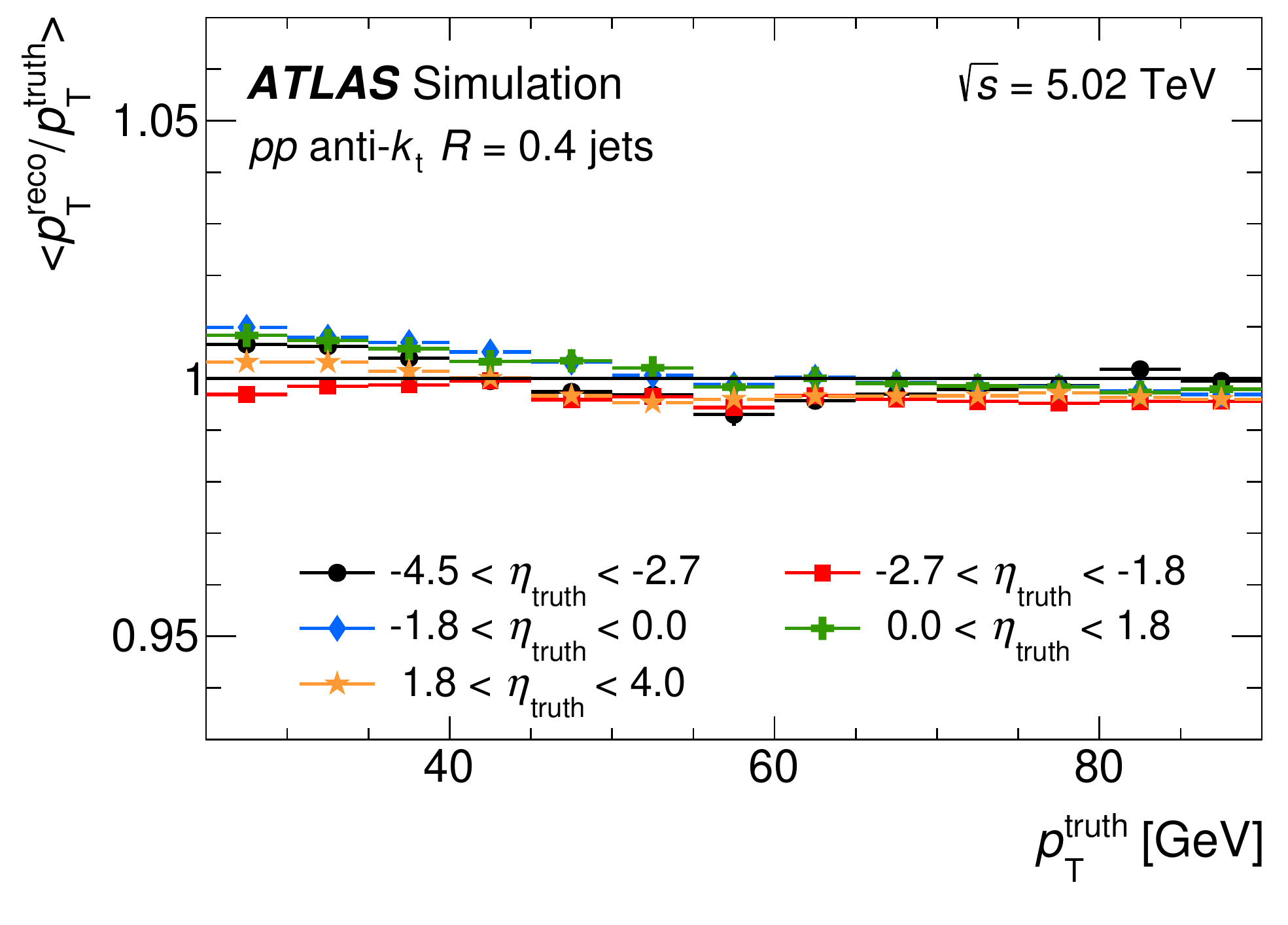}  &
			\includegraphics[width=0.45\textwidth]{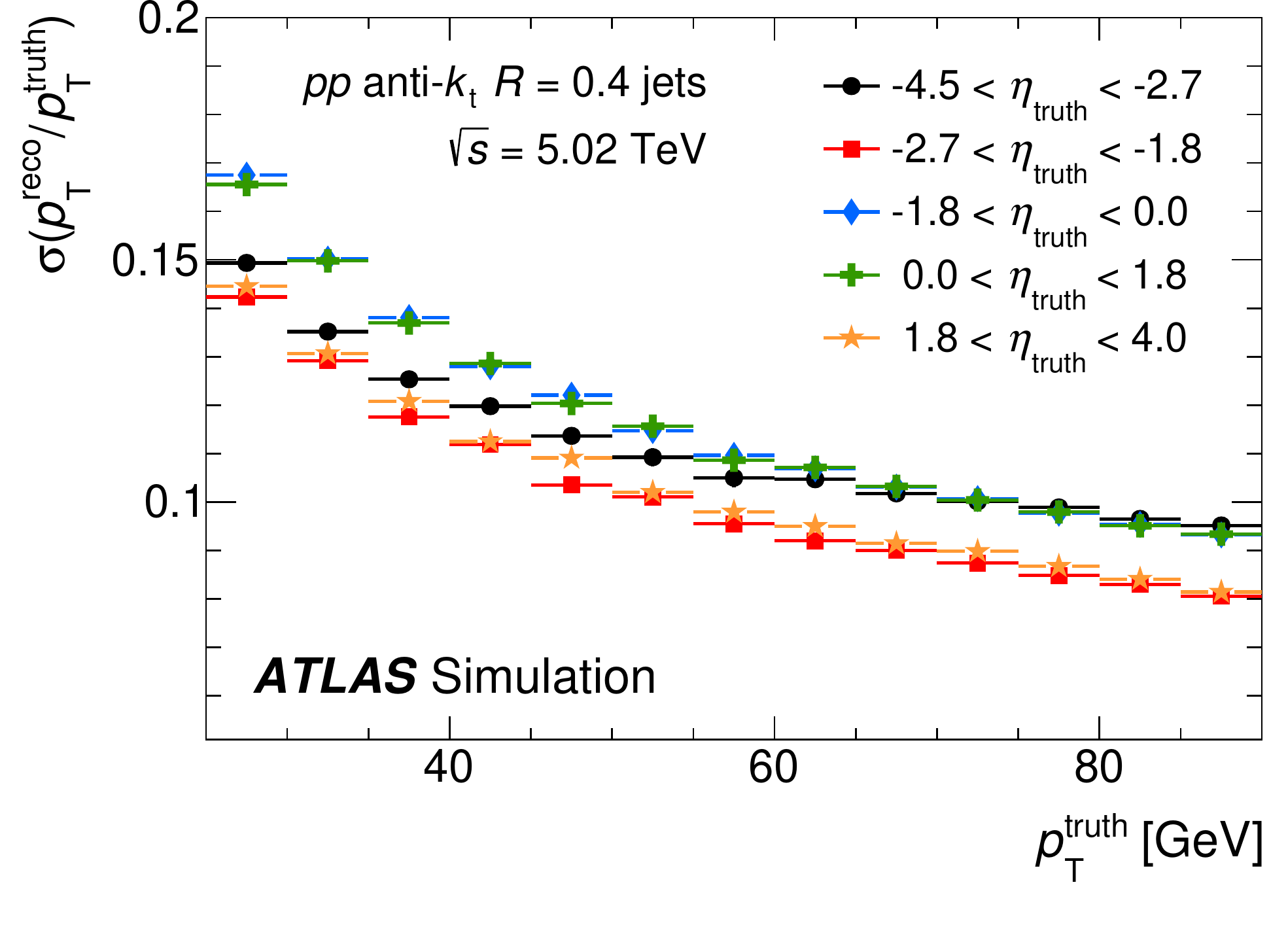} \\	
			\includegraphics[width=0.45\textwidth]{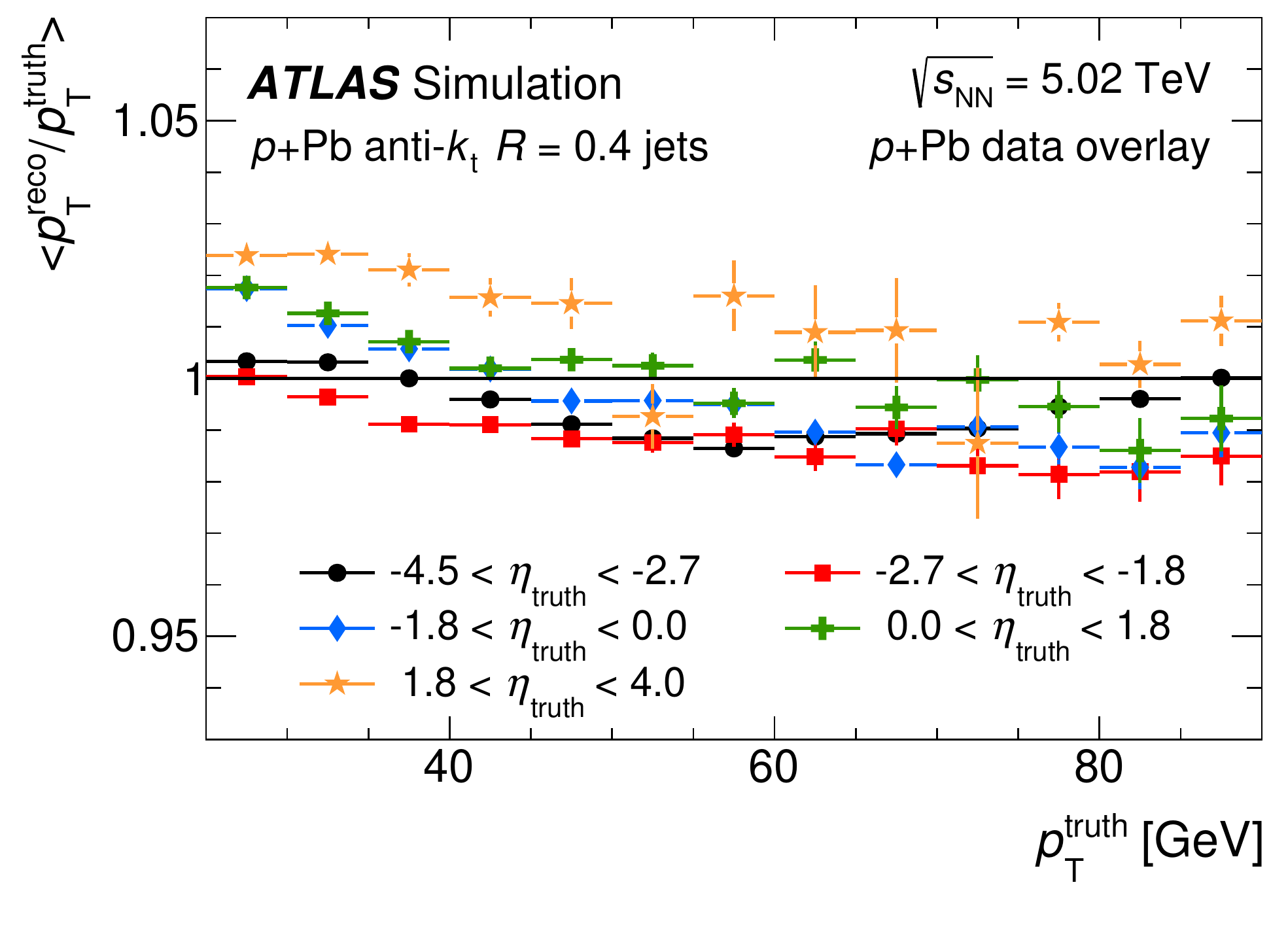} &
			\includegraphics[width=0.45\textwidth]{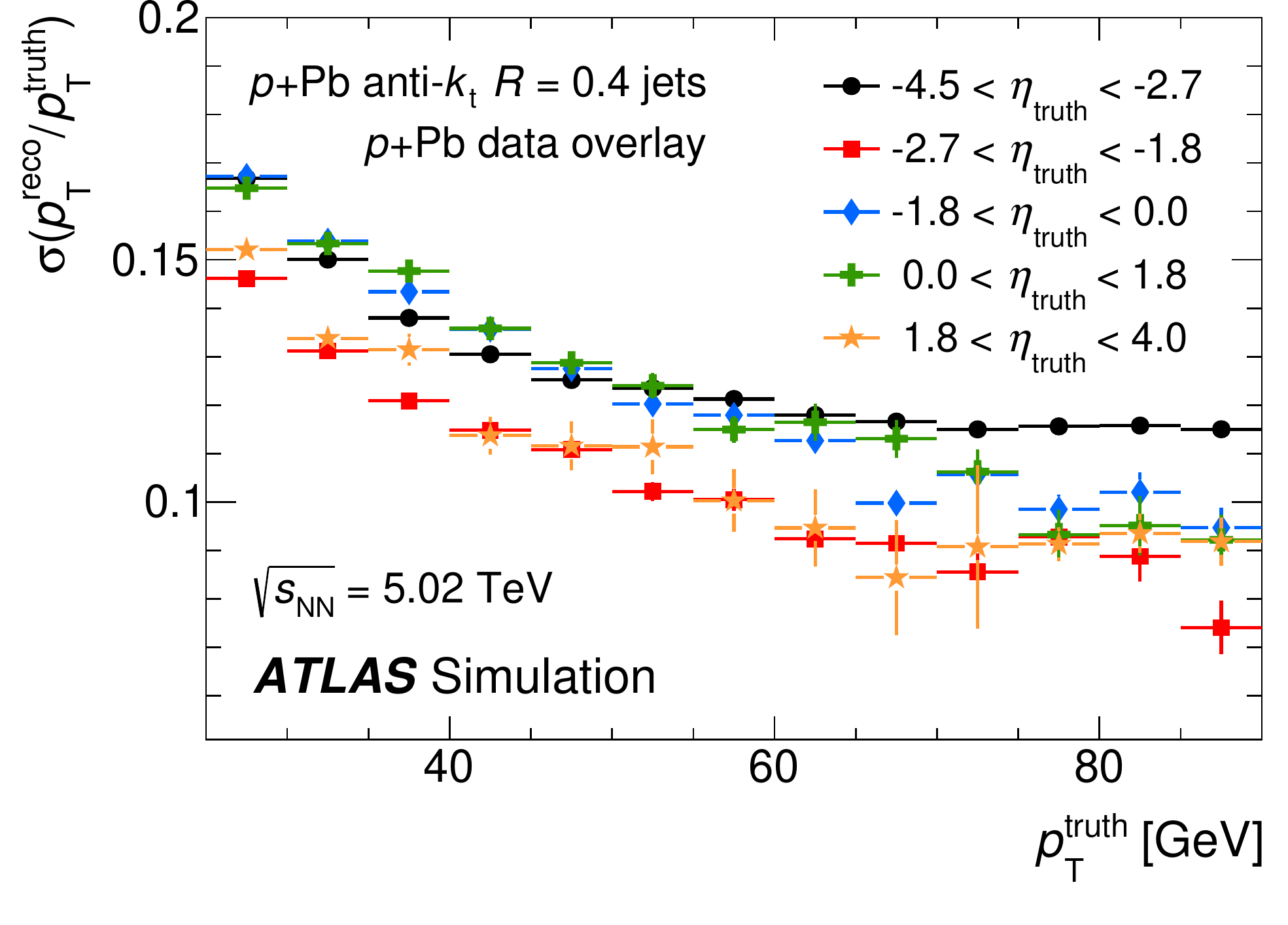} \\	\end{tabular}		
	}
	\caption{(Left) Jet energy scale and (right) jet energy resolution evaluated in (top) \pp\ and (bottom) \pPb\ MC samples in different generator-level jet pseudorapidity intervals and shown as a function of the generator-level jet transverse momentum \pttruth. }
	\label{fig:jesjerpp}
\end{figure}

The jet energy reconstruction performance is characterized using the ratios of transverse momenta of reconstructed jets to generated jets, \ptreco\ and \pttruth\  respectively, to determine the relevant jet energy scale~(JES), and jet energy resolution~(JER) corresponding to the mean and width of the jet response $(\ptreco/\pttruth)$. The values of JES and JER are shown in Figure~\ref{fig:jesjerpp} as a function of \pttruth, in intervals of generated jet pseudorapidity \etatruth, for \pp\ and \pPb\ MC samples. The JES shows a very small dependence on \etatruth, with a maximum deviation of $\pm 3\%$ from unity. Jet angular reconstruction performance has been studied  in terms of mean angular differences between the reconstructed and generator-level jet direction in pseudorapidity and azimuthal angle, $\langle\Deta\rangle$ and $\langle\Dphi\rangle$, and their resolutions $\sigma(\Deta)$ and $\sigma(\Dphi)$. The mean angular differences are consistent with zero, and the jet angular resolutions (JAR) decrease from approximately 17\% to 10\% as a function of \pttruth\ for both the \pp\ and \pPb\ MC samples.

\section{Analysis procedure}
\label{sec:analysis}
The two-highest \pt\ jets in each event are used to measure the azimuthal angular correlation distributions, which are evaluated as a function of \Dphi\ relative to the leading jet in the center-of-mass rapidity interval $2.7<\ystarone<4.0$, and in different intervals of \ystartwo, \ptone, and \pttwo. Table~\ref{tab:binning} lists the transverse momenta and center-of-mass rapidity intervals used in the measurement. The \conetwo\ distributions are then fitted to extract their widths.

\begin{table}[h]
	\caption{\label{tab:binning} The transverse momentum intervals (\ptone, \pttwo) of the leading and subleading jets and the center-of-mass rapidity intervals (\ystartwo) of the subleading jet. In all cases the center-of-mass rapidity interval of the leading jet is $2.7<\ystarone<4.0$. }
	\centering
	\begin{tabular}{|| c | c | c || } 
		\hline
		Bins in \ptone\ [\GeV] & Bins in \pttwo\ [\GeV] & Bins in \ystartwo\\ 
		\hline
		$28<\ptone<35$   & $28<\pttwo<35$  & $\ \  2.7<\ystartwo<4.0$ \\ 
		$35<\ptone<45$   & $35<\pttwo<45$  & $\ \  1.8<\ystartwo<2.7$ \\ 
		$45<\ptone<90$   & $45<\pttwo<90$  & $\ \  0.0<\ystartwo<1.8$ \\
		& 				   & $-1.8<\ystartwo<0.0$ \\
		&				   & $\ \ \ -4.0<\ystartwo<-1.8$ \\
		\hline
	\end{tabular}
\end{table}

The effects of migration due to the jet energy and angular resolutions as well as the jet reconstruction efficiency affecting the leading-jet \pT\ spectra and \conetwo\ distributions in \pp\ and \pPb\ collisions are corrected for by using a bin-by-bin unfolding procedure. For each of the affected distributions, correction factors that are  applied to data are derived from the ratio between two corresponding MC distributions; one evaluated using generator-level jets and the other evaluated using jets reconstructed after the detector simulation. To account for the jets excluded due to the disabled HEC region in \pPb\ data and MC samples, an acceptance correction is applied using the same procedure because generator-level jets are not excluded from the affected region. Thus, the correction factors used in the unfolding account for the missing jets at reconstruction level. The bin-by-bin unfolding procedure is sensitive to differences in the shapes of distributions between the data and the MC samples. Thus, the jet \pT\ and \conetwo\ distributions in the  MC reconstructed samples are reweighted to match the shapes in the data. Weights are derived by evaluating the data-to-MC ratios of the reconstructed distributions. The reweighting is done in two steps: 1) weights are evaluated for the jet \pt\ spectra; 2) when deriving weights for the \conetwo\ distributions, the dependence of the ratio between data and MC on the jet \pt\ spectra is removed by applying the weights evaluated in the previous step. The final weight is the product of the two weights. Jet weights of the jet \pt\ spectra are within 10\% of unity for \pp\ and \pPb\ collisions, and the \Dphi\ weights are within 15\% of unity near the peak of the \conetwo\ distributions, where the effect of reweighting is largest. 

The unfolded jet \pT\ and $dN_{12}/d\Dphi$ distributions are used to evaluate the \conetwo\ distributions both in \pp\ and in \pPb\ collisions. The \conetwo\ distributions are then fitted as a function of $\Delta\Phi = \Dphi - \pi$ by a symmetric exponential distribution convolved with a Gaussian function:
\begin{eqnarray}
\conetwo(\Delta\phi) = \int_{-\infty}^{\infty}d\delta\frac{\mathrm{e}^{-\delta^{2}/2\sigma^{2}}}{\sqrt{8\pi\sigma^{2}\tau^{2}}}\mathrm{e}^{-|\Delta\Phi -\delta|/\tau}, \nonumber
\end{eqnarray}
where $\tau$ is the parameter of the exponential component and $\sigma$ is the width of the Gaussian distribution. All parameters are required to be positive. The resulting fit function is:
\begin{eqnarray}
\conetwo(\Delta\phi) = A\frac{\mathrm{e}^{\sigma^2/2\tau^2}}{2\tau}\bigg(\frac{1}{2}\mathrm{e}^{\Delta\Phi/ \tau}\mathrm{Erfc}\bigg(\frac{1}{\sqrt{2}}\bigg[\frac{\Delta\Phi}{\sigma}+\frac{\sigma}{\tau}\bigg]\bigg)+\mathrm{e}^{-\Delta\Phi/\tau}\bigg[1-\frac{1}{2}\mathrm{Erfc}\bigg(\frac{1}{\sqrt{2}}\bigg[\frac{\Delta\Phi}{\sigma}-\frac{\sigma}{\tau}\bigg]\bigg)\bigg]\bigg), \nonumber
\end{eqnarray} 
where $A$ is a normalization factor. The width \wonetwo\ is chosen to be represented by the analytic root-mean-square of the $\tau$ and $\sigma$ parameters resulting from the fit, $\wonetwo = \mathrm{RMS} (\conetwo) =  \sqrt{2\tau^2 + \sigma^{2}}$. The fitting procedure is performed in the range $2.5<\Dphi<\pi$. The convolution of the Gaussian and symmetric exponential functions is found to better describe the data around the peak of the \conetwo\ distributions than a pure exponential function.

\section{Systematic uncertainties}
\label{sec:systematics}
Systematic uncertainties originate from the JES, JER, JAR, the fitting procedure, acceptance correction, and unfolding procedure. For each source of systematic uncertainty, the values of \wonetwo\ and \ionetwo\ and the ratios \cppb\ and \ippb\ in \pPb\ and \pp\ collisions are re-evaluated. The absolute difference between the varied and nominal values is used as an estimate of the uncertainty.

The systematic uncertainty due to the JES is determined from in situ studies of the calorimeter response~\cite{Aad:2011he,HIjesnote,Aaboud:2017jcu,Aaboud:2018twu}, and studies of a relative energy-scale difference between the heavy-ion jet reconstruction procedure~\cite{HIjesnote} and the procedure used in 13~\TeV\ \pp\ collisions~\cite{Aad:2014bia}. The JES uncertainty depends on the jet \pT\ and jet $\eta$ and is applied as a modification to the reconstructed jet \pT\ and varied separately by $\pm$1 standard deviation. The bin-by-bin correction factors are recomputed accordingly and the data are unfolded with them. The resulting uncertainty from the JES is typically less than 15\% for the values of both \wonetwo\ and \ionetwo. An additional source of systematic uncertainty for the JES in \pPb\ collisions originates from differences between detector response and its simulation compared to \pp\ collisions. These differences are about 1\%, and their resulting systematic uncertainties are added to the total JES systematic uncertainty in quadrature.

The uncertainty due to the JER is evaluated by repeating the unfolding procedure with modified bin-by-bin correction factors, where an additional contribution is added to the resolution of the simulated jet \pt\ using a Gaussian smearing procedure~\cite{Aad:2014bia}. The smearing factor is evaluated with an in situ technique developed for 13~\TeV\ \pp\ data involving studies of dijet transverse momentum balance~\cite{Aaboud:2018kfi}. An additional uncertainty is included to account for differences between the heavy-ion jet reconstruction and that used in the analyses of 13~\TeV\ \pp\ data. The resulting uncertainty is symmetrized. The size of the uncertainty due to the JER for the values of \ionetwo\ is as large as 30\% and is typically below 10\% for the values of \wonetwo. 

The systematic uncertainty from the JAR originates in differences in the angular resolution between the data and MC samples. The uncertainty is derived as the difference between the angular resolutions evaluated using the two different MC generators, \herwig\ and \pythiaeight. Distributions are unfolded with modified bin-by-bin correction factors where the reconstructed jet $\eta$ and $\phi$ are smeared to reflect an up to $\sim$5\% uncertainty of the JAR. The size of the resulting uncertainty on \wonetwo\ and \ionetwo\ is typically below 6\%.

A systematic uncertainty related to a possible dependence of the result on the fit range is considered. This systematic uncertainty is present only for the values of \wonetwo\ and \cppb. The uncertainty is evaluated by modifying the fit interval from the default of $2.5<\Dphi<\pi$ to a fit range of $2.1<\Dphi<\pi$. In different ranges of \ptone\ and \pttwo, the resulting uncertainties are fitted to a constant function over the range $|\ystar|<4.0$. The systematic uncertainty is smoothed by a fit in order to minimize the impact of the statistical fluctuations. The size of the resulting uncertainty of \wonetwo\ is less than $7\%$. 

The systematic uncertainty from the bin-by-bin unfolding procedure is associated with  differences in the shapes of distributions between the data and MC samples. To achieve better correspondence with the data, the simulated values are reweighted to match the shapes in the data. The entire change in the unfolded values induced by the use of reweighted bin-by-bin correction factors is taken as the systematic uncertainty, which is below 5\% for \conetwo\ and \ionetwo. 

The systematic uncertainty associated with the acceptance correction for the disabled part of the HEC during \pPb\ data taking is evaluated by increasing the size of the excluded region by 0.1 in azimuth and pseudorapidity, which corresponds to the size of the calorimeter towers. The resulting uncertainty is symmetrized to account for no reduction in the size of the excluded region due to the simultaneous overlap of the jet area with the regions covered by the enabled and disabled HEC. The uncertainty only affects the rapidity region $-4.0<\ystartwo<-1.4$. The resulting uncertainty of \wonetwo\ is negligible. The yields \ionetwo\ have an uncertainty of up to 10\%.

\begin{figure}[htbp]
	\centerline{
		\begin{tabular}{cc}
			\includegraphics[width=0.5\textwidth]{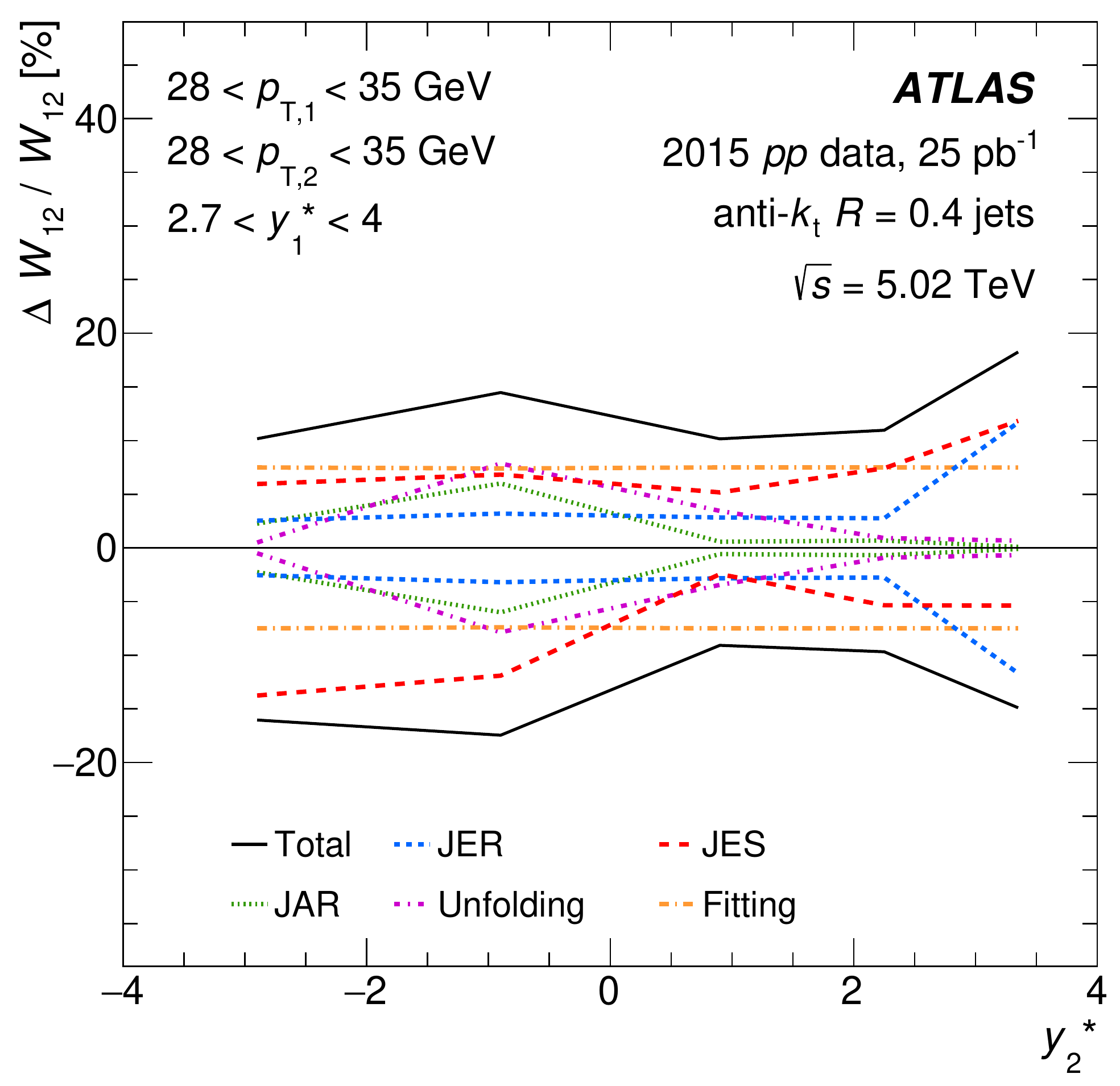} &
			\includegraphics[width=0.5\textwidth]{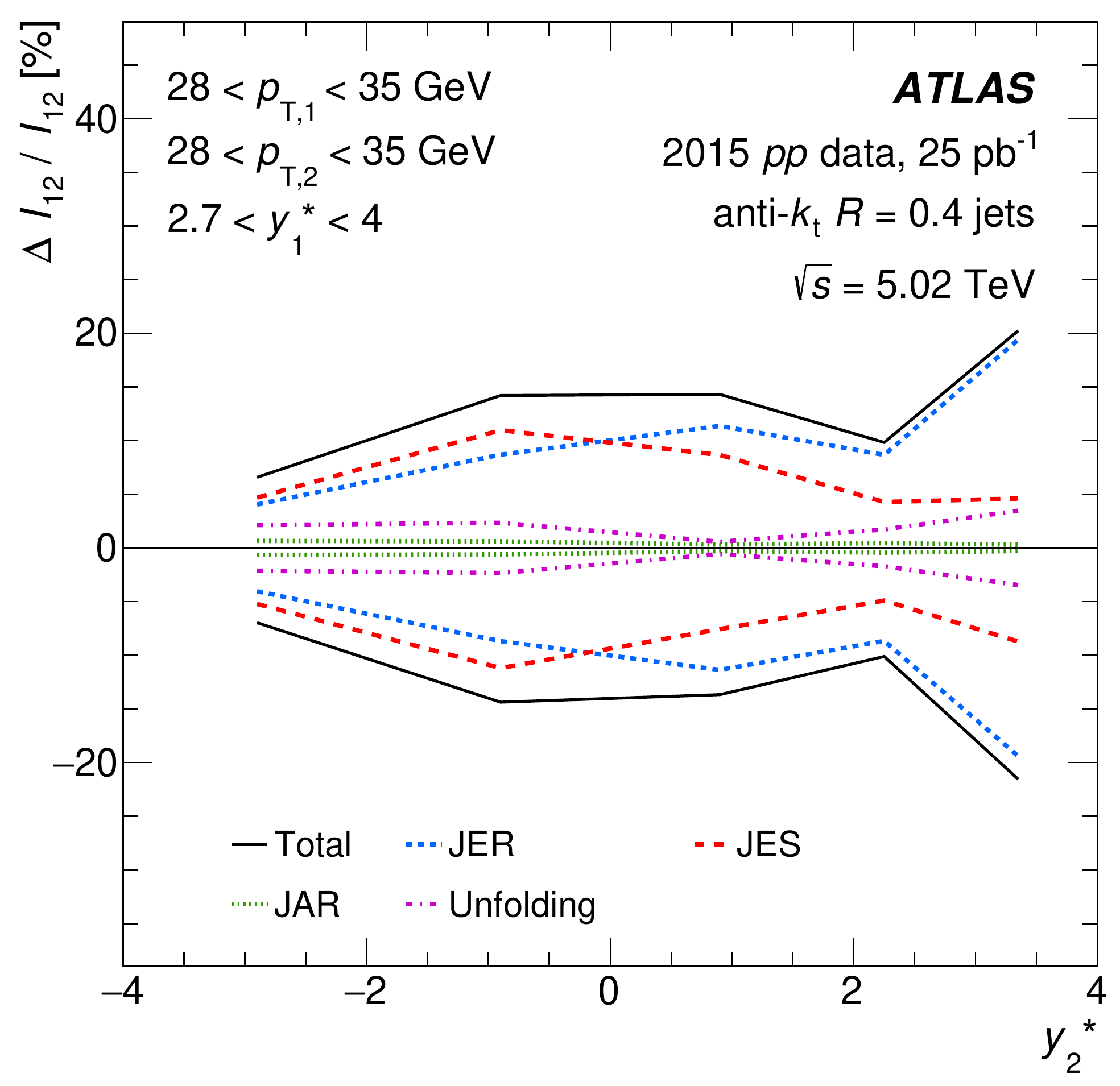} \\
			\includegraphics[width=0.5\textwidth]{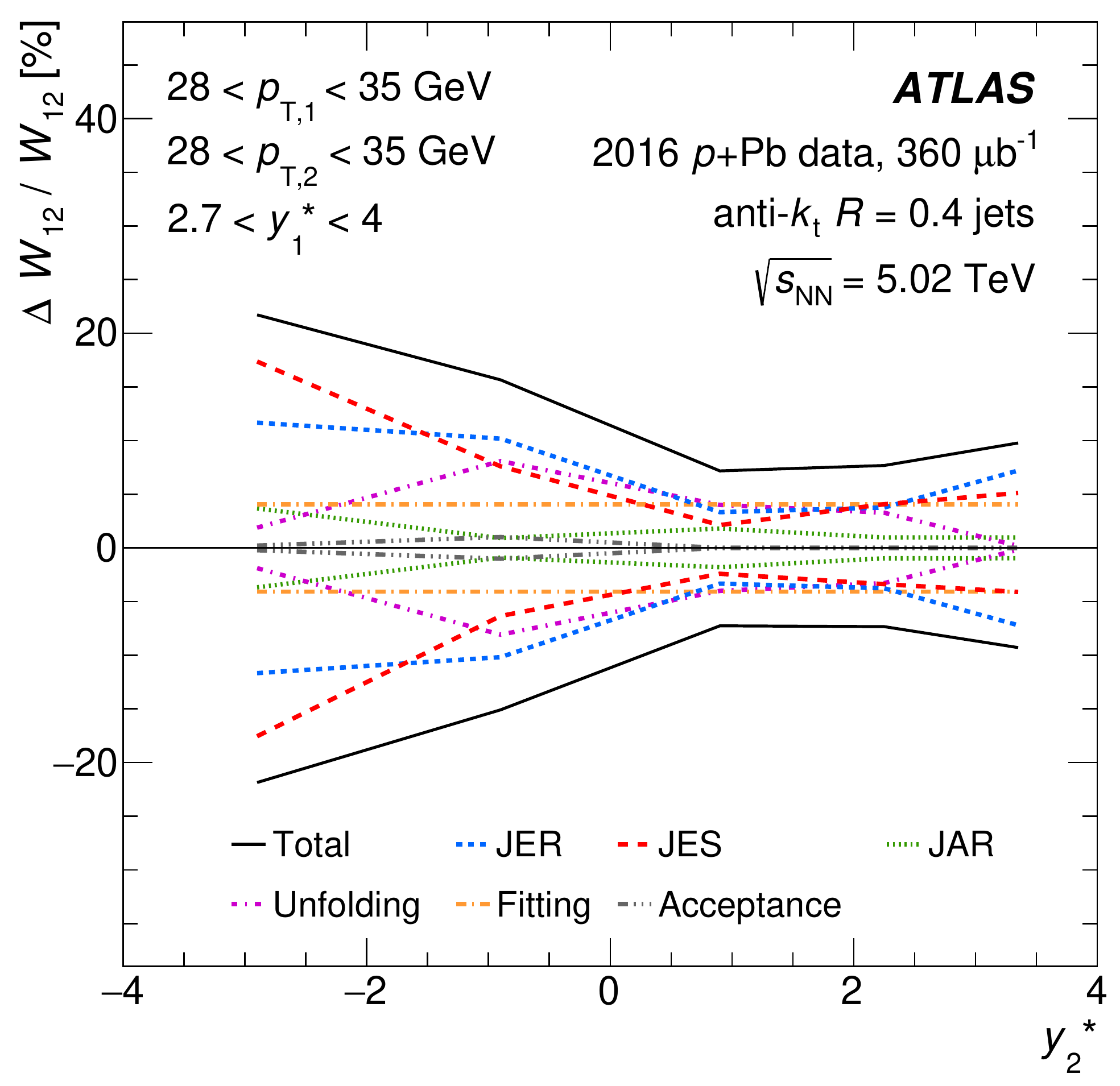} &
			\includegraphics[width=0.5\textwidth]{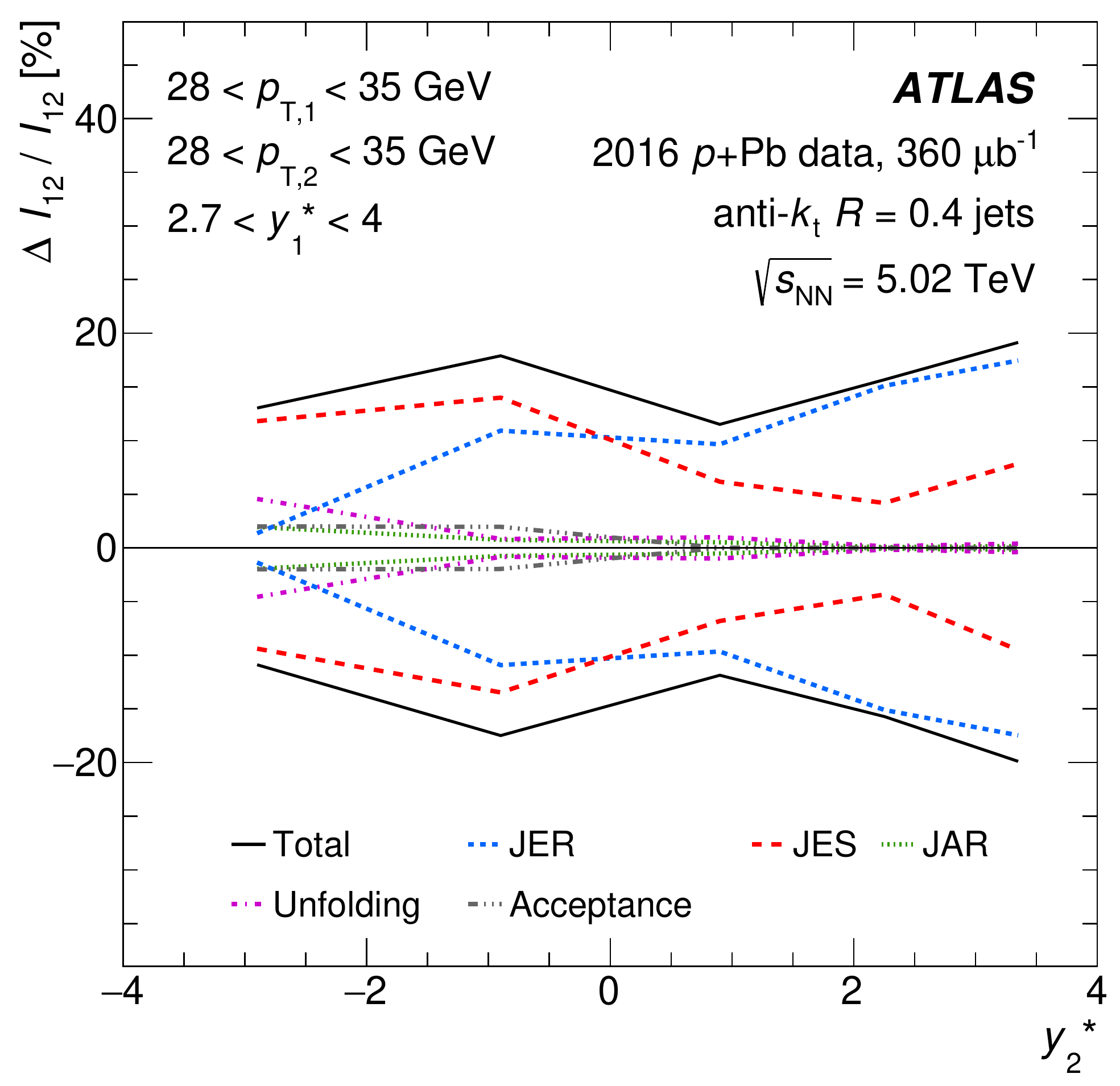} \\
		\end{tabular}
	}
	\caption{Relative systematic uncertainties of values of (left) \wonetwo\ and (right) \ionetwo\ in (top) \pp\ and (bottom) \pPb\ collisions. The uncertainty associated with the disabled HEC region is labeled as the ``Acceptance'' uncertainty. Uncertainty values are presented for the center of the bin and with no $\Delta\pt$ requirement.}
	\label{fig:distsys}
\end{figure}

For these measurements, the systematic uncertainties in the values of \wonetwo\ and \ionetwo\ are presented in Figure~\ref{fig:distsys}. The systematic uncertainties from each source are assumed to be uncorrelated and are thus combined in quadrature to obtain the total systematic uncertainty.  

\begin{figure}[htbp]
	\centerline{
		\begin{tabular}{cc}
			\includegraphics[width=0.5\textwidth]{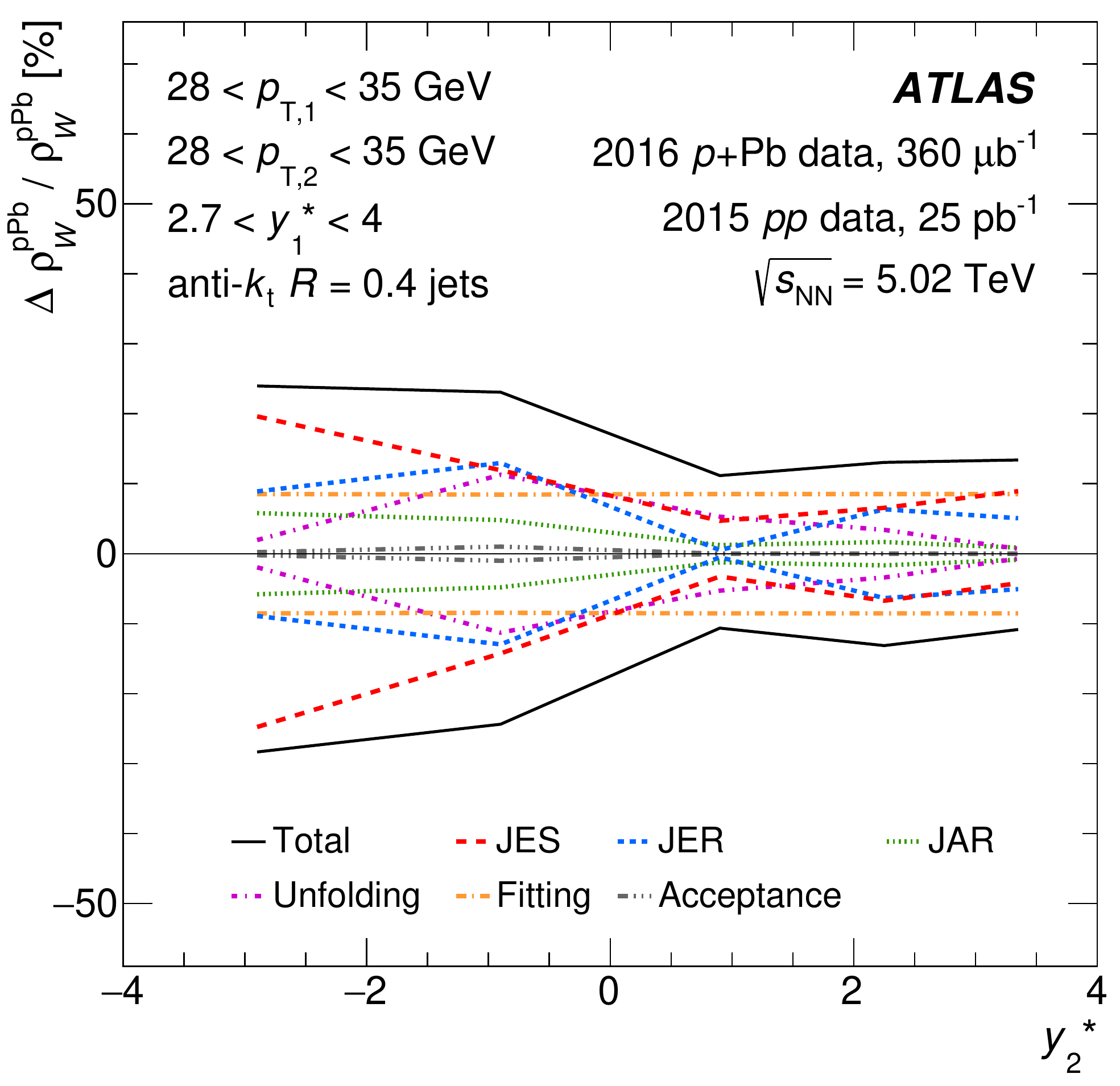} &
			\includegraphics[width=0.5\textwidth]{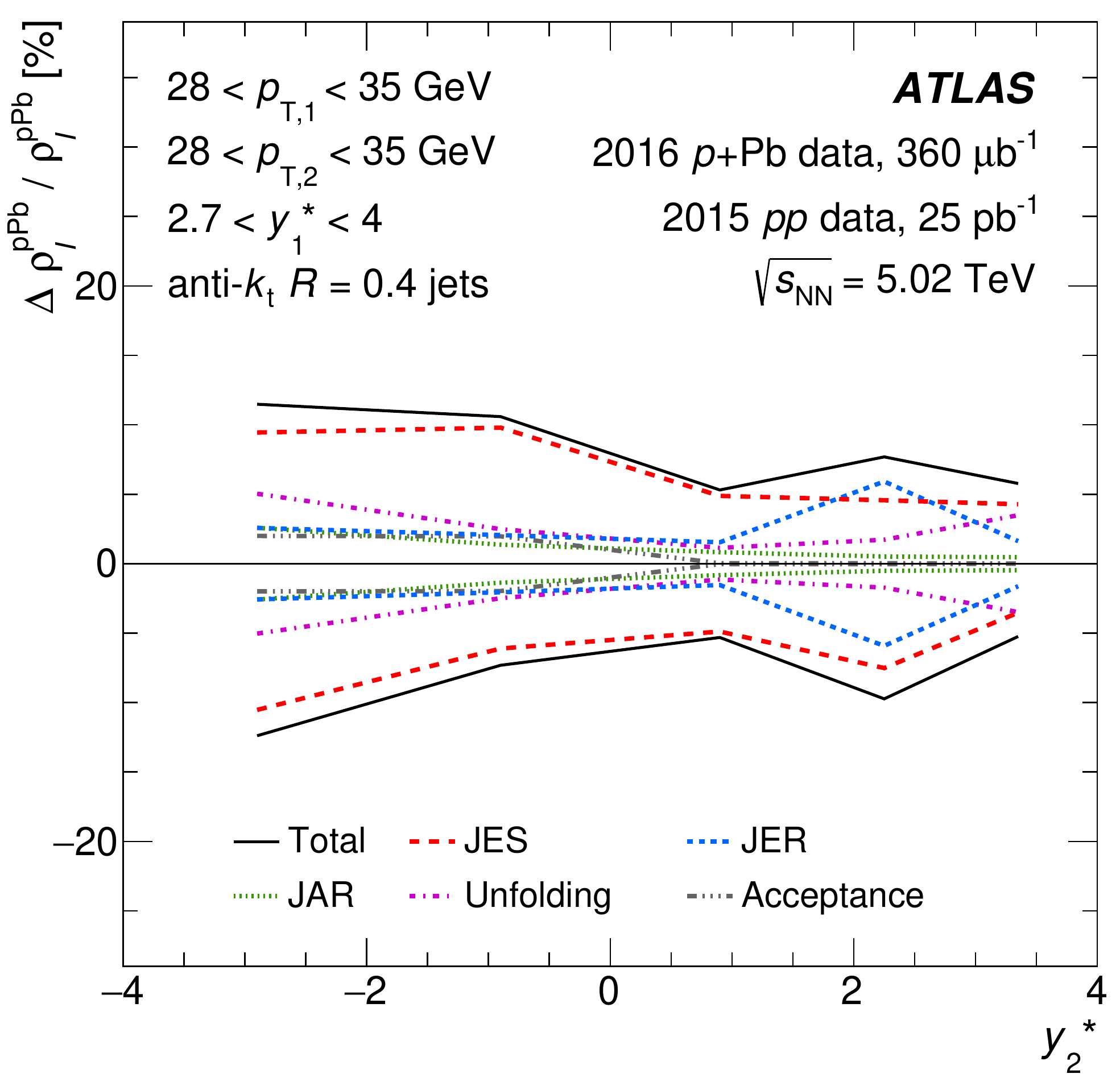} \\
		\end{tabular}
	}
	\caption{Relative systematic uncertainties of values of (left) \cppb\ and (right) \ippb . The uncertainty associated with the disabled HEC region is labeled as the ``Acceptance'' uncertainty. Uncertainty values are presented for the center of the bin and with no $\Delta\pt$ requirement. }
	\label{fig:ratiosyst}
\end{figure}

In evaluating the \pPb\ to \pp\ ratios, the correlations between the various systematic uncertainties are considered. The uncertainties associated with unfolding, fitting, the acceptance correction, and the additional JES uncertainties associated with the differences between the detector response and its simulations in \pPb\ collisions compared to \pp\ collisions are taken to be uncorrelated between the two collision systems and are added in quadrature. All other uncertainties associated with the JES, JER, and JAR are taken to be correlated. To account for correlations, the ratios are re-evaluated by applying variations to both collision systems simultaneously. The resulting variations of the ratios from their central values are used as the correlated systematic uncertainty from a given source. Examples of systematic uncertainties for the values of \cppb\ and \ippb\  are presented in Figure~\ref{fig:ratiosyst}, where the systematic uncertainty from the JES (up to 20\%) is dominant.

\FloatBarrier

\section{Results}
\label{sec:results}
\begin{figure}[hb]
	\centerline{
		\begin{tabular}{ccc}
			\includegraphics[width=0.47\textwidth]{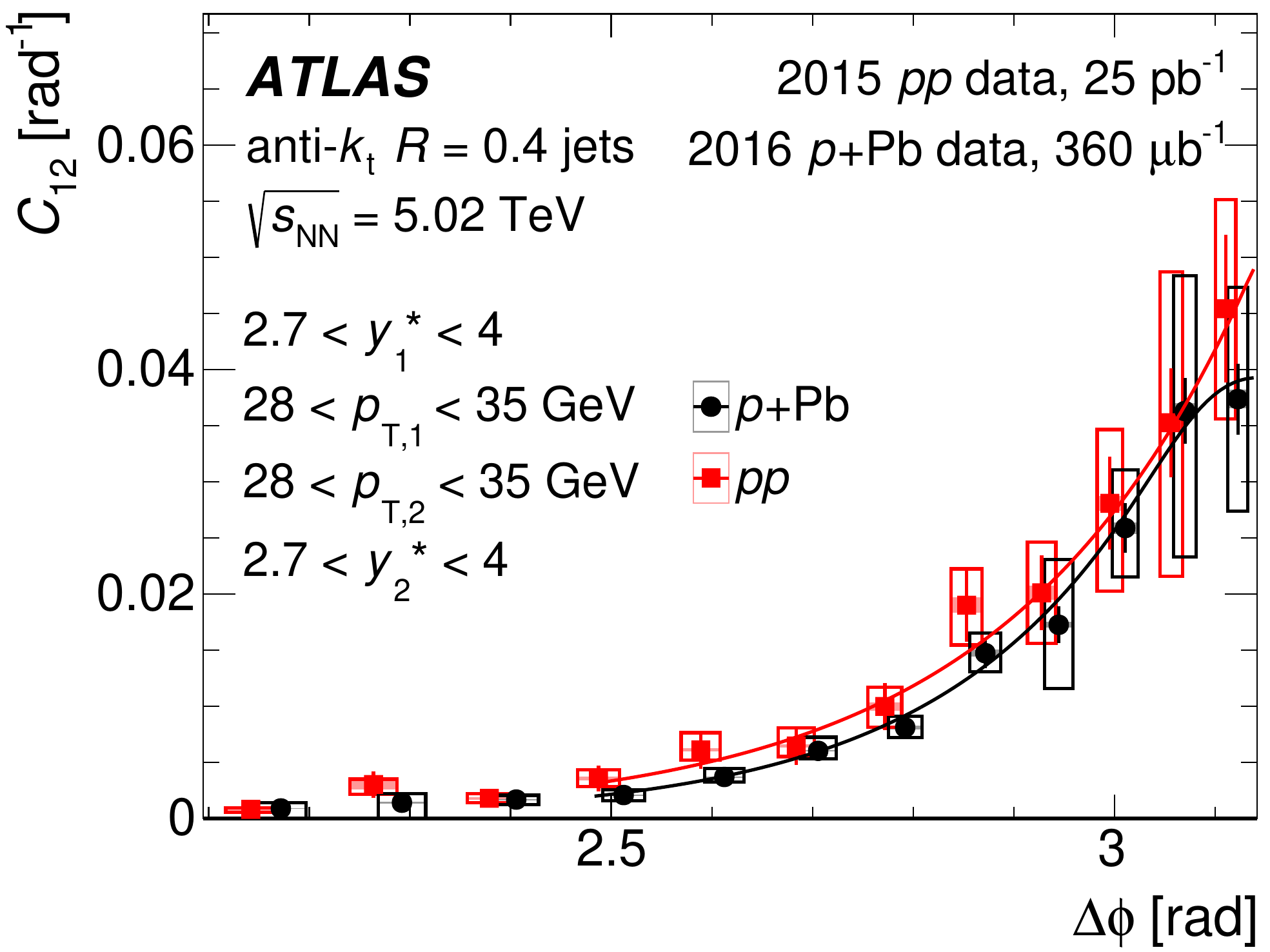} & 
			\includegraphics[width=0.47\textwidth]{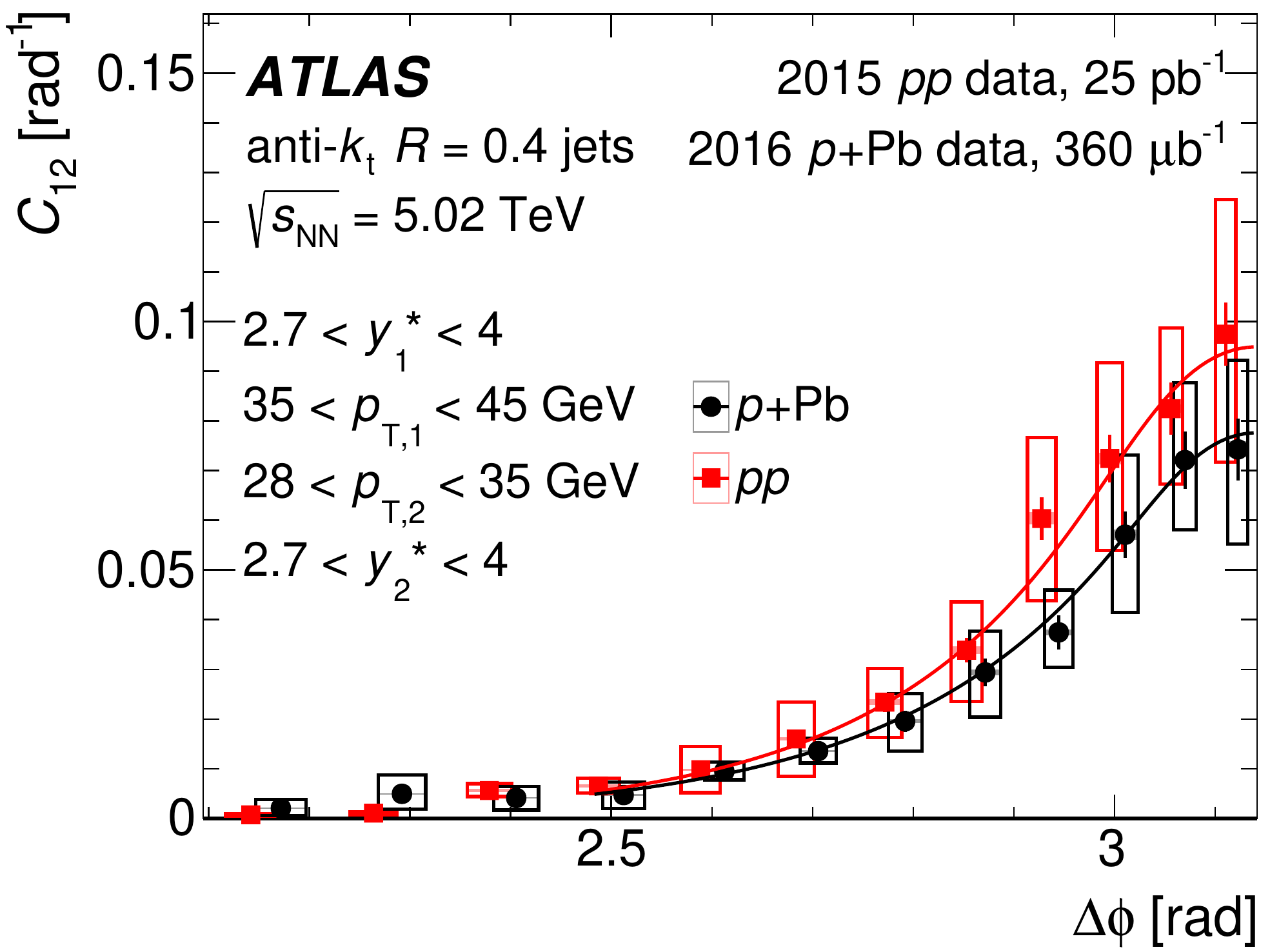} \\ 
			\includegraphics[width=0.47\textwidth]{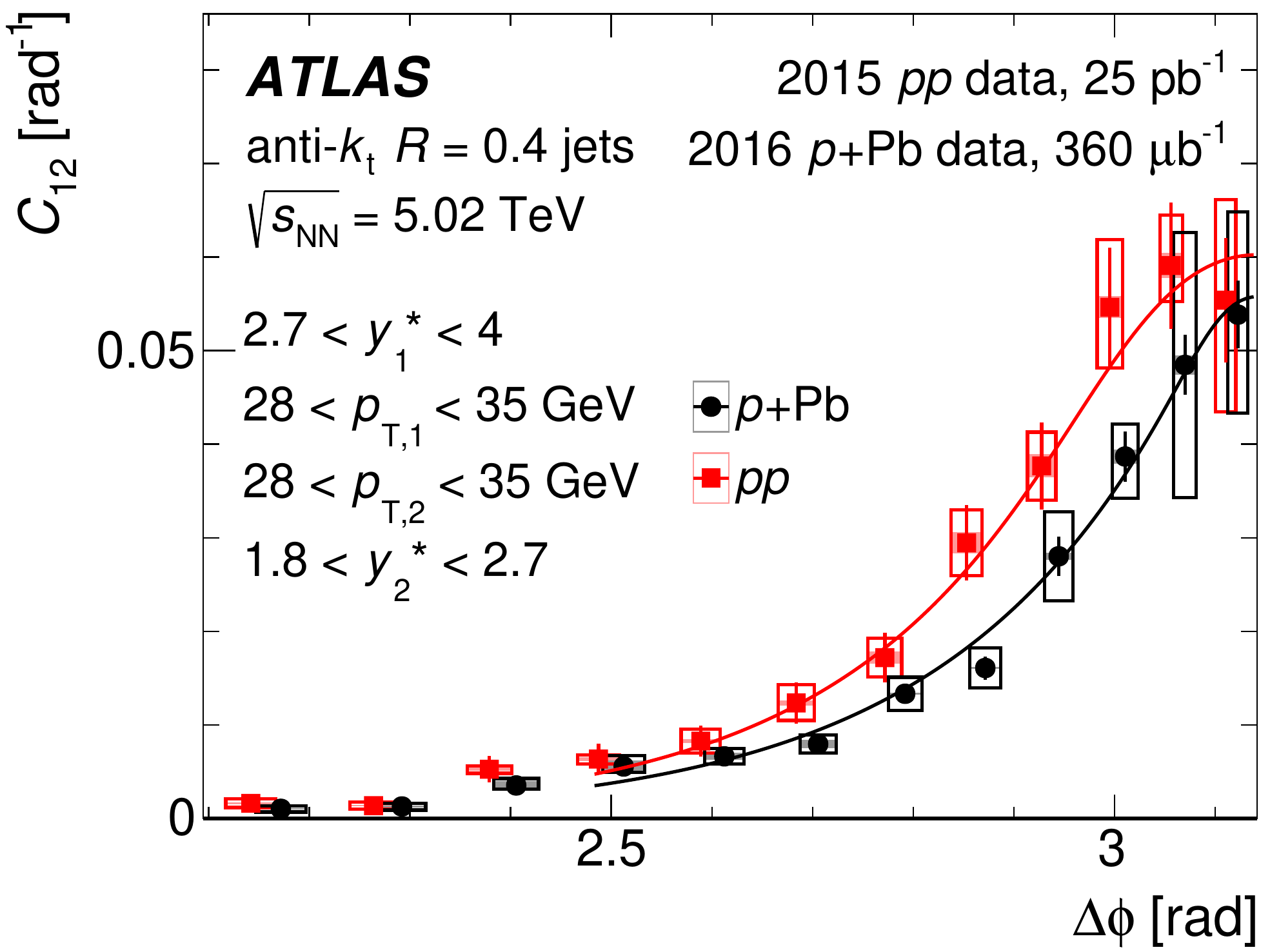} &
			\includegraphics[width=0.47\textwidth]{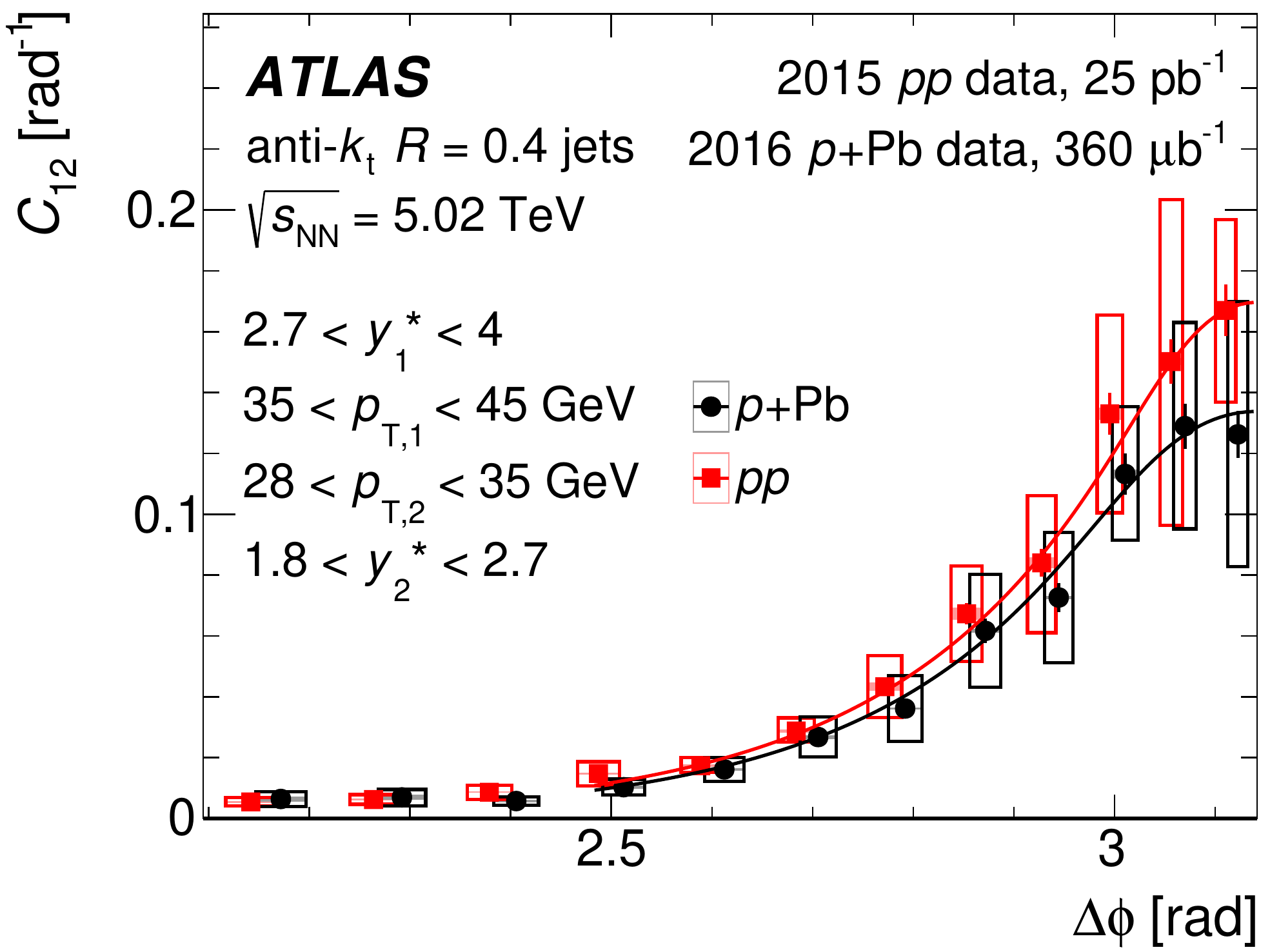} \\
		\end{tabular}
	}
	\caption{ Unfolded \conetwo\ distributions in (red squares) \pp\  and (black circles) \pPb\ collisions for different selections of \ptone, \pttwo, \ystarone, and \ystartwo\ as a function of \Dphi. The lines represent values of the fit function. The data points are shifted horizontally for visibility, and do not reflect an actual shift in \Dphi. The vertical size of the open boxes represents systematic uncertainties and error bars indicate statistical uncertainties. The horizontal size of the open boxes does not represent the width of the bins. Results are shown with no $\Delta\pt$ requirement, where $\Delta\pt=\ptone-\pttwo$. }
	\label{fig:dphipPbpp}
\end{figure}

This section presents values of \wonetwo\ and \ionetwo\ and the ratios \cppb\ and \ippb\ in \pPb\ and \pp\ collisions. Examples of unfolded \conetwo\ distributions in different intervals of  \ptone and \pttwo\ evaluated in \pp\ and \pPb\ collisions are shown in Figure~\ref{fig:dphipPbpp} together with the fit results. The \conetwo\ distributions have a characteristic peak at $\Dphi = \pi$.

\begin{figure}[hb]
	\centerline{
		\begin{tabular}{cc}
			\includegraphics[width=0.5\textwidth]{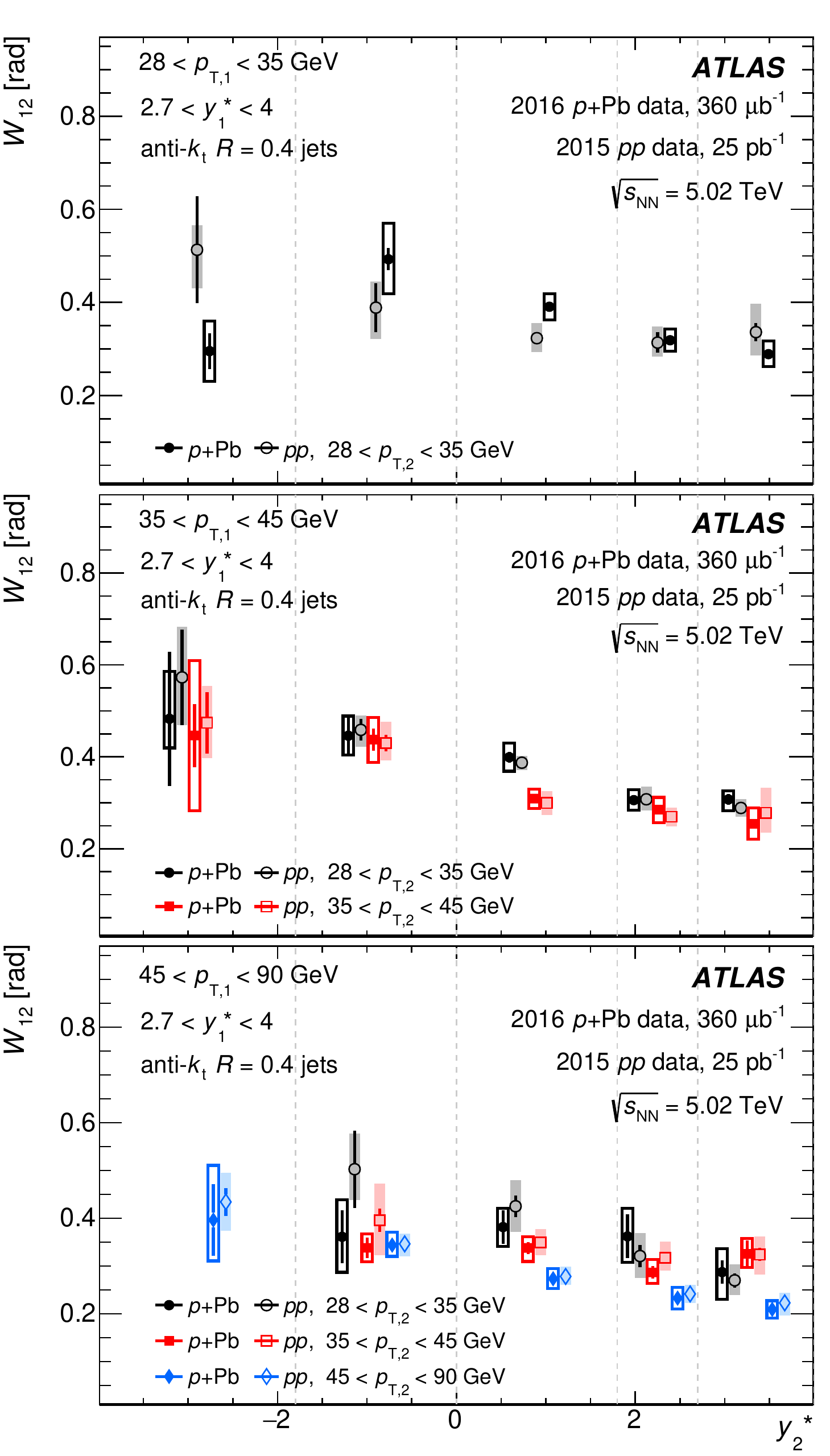} &
			\includegraphics[width=0.5\textwidth]{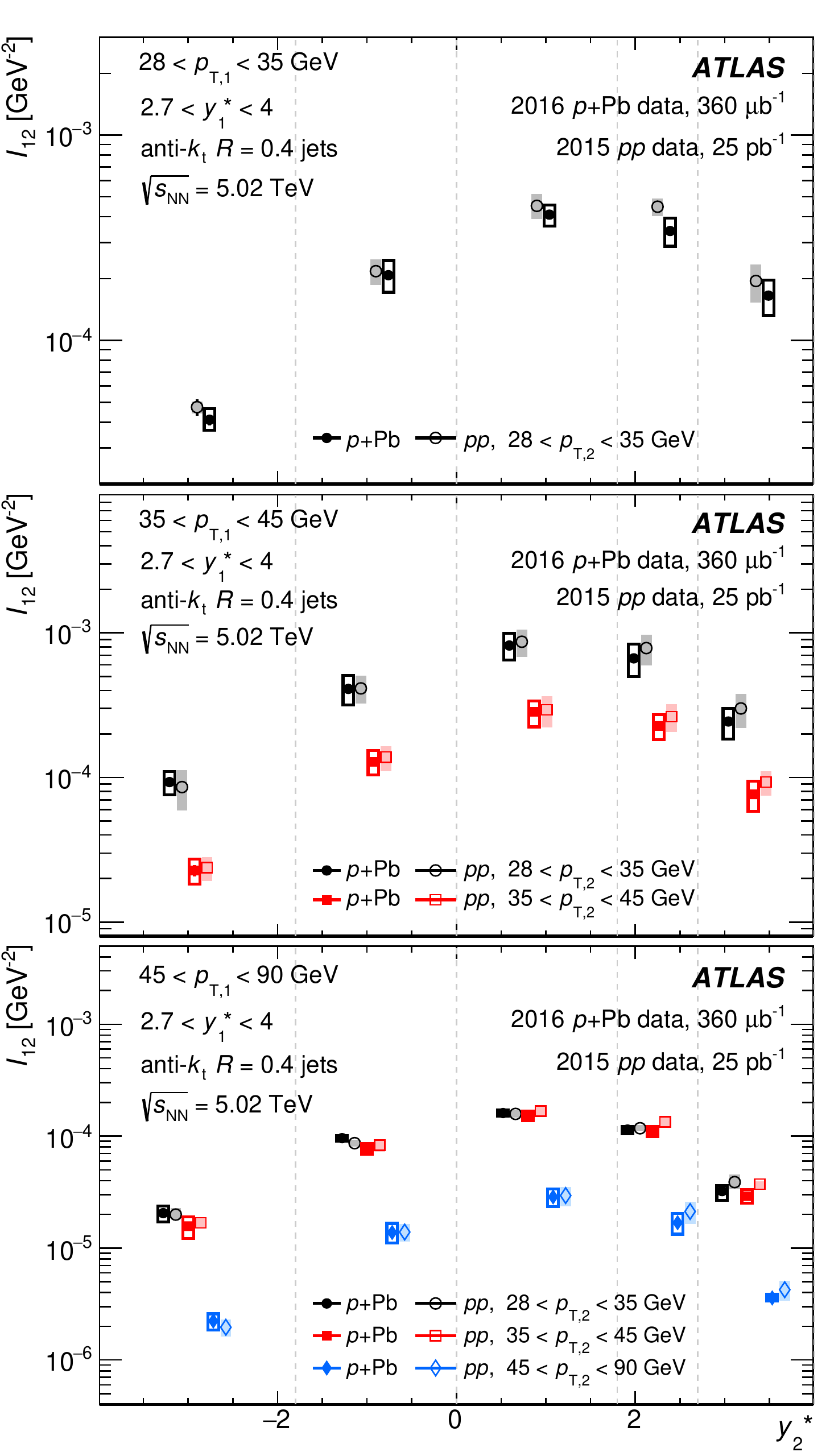} \\
		\end{tabular}
	}
	\caption{Comparison of (left) \wonetwo\ and (right) \ionetwo\ values in \pp\ (open symbols) and \pPb\ (closed symbols) collisions for different selections of \ptone\ and \pttwo\ as a function of \ystartwo. The \ystartwo\ intervals are separated by dotted vertical lines. The data points are shifted horizontally for visibility, and do not reflect an actual shift in rapidity. The vertical size of the shaded and open boxes represents systematic uncertainties for \pp\ and \pPb, respectively, and the error bars indicate statistical uncertainties. The horizontal size of the shaded and open boxes does not represent the width of the bins. Some points are not presented due to large statistical uncertainties. Results are shown with no $\Delta\pt$ requirement, where $\Delta\pt=\ptone-\pttwo$. }
	\label{fig:finalplots}
\end{figure}

The results of measurements of \wonetwo\ in \pPb\ and \pp\ collisions for different ranges of \ptone\ and \pttwo\ as a function of \ystartwo\ are  presented in left panels of Figure~\ref{fig:finalplots}. The value of \wonetwo\ decreases with decreasing rapidity separation $(|\ystarone - \ystartwo|)$ between the leading and subleading jets in both the \pp\ and \pPb\ collisions. The value of \wonetwo\ increases with imbalance in \pT\ between the leading and subleading jets.  The results of the measurement of conditional yields \ionetwo\ in \pPb\ and \pp\ collisions are shown in the right panels of Figure~\ref{fig:finalplots}. Initially, the value of \ionetwo\ increases with decreasing separation in rapidity between the two jets, reaching a maximum for subleading jets in the interval $0.0<\ystartwo<1.8$, and then decreases for smaller rapidity separations between the two jets. This is attributed to the decrease of the dijet cross-section at large rapidity being faster than that of the inclusive jet cross-section. The distributions of \ionetwo\ have similar shapes in \pp\ and \pPb\ collisions for all \ptone\ and \pttwo\ combinations.

\begin{figure}[ht]
	\centerline{
		\begin{tabular}{cc}
			\includegraphics[width=1.2\textwidth]{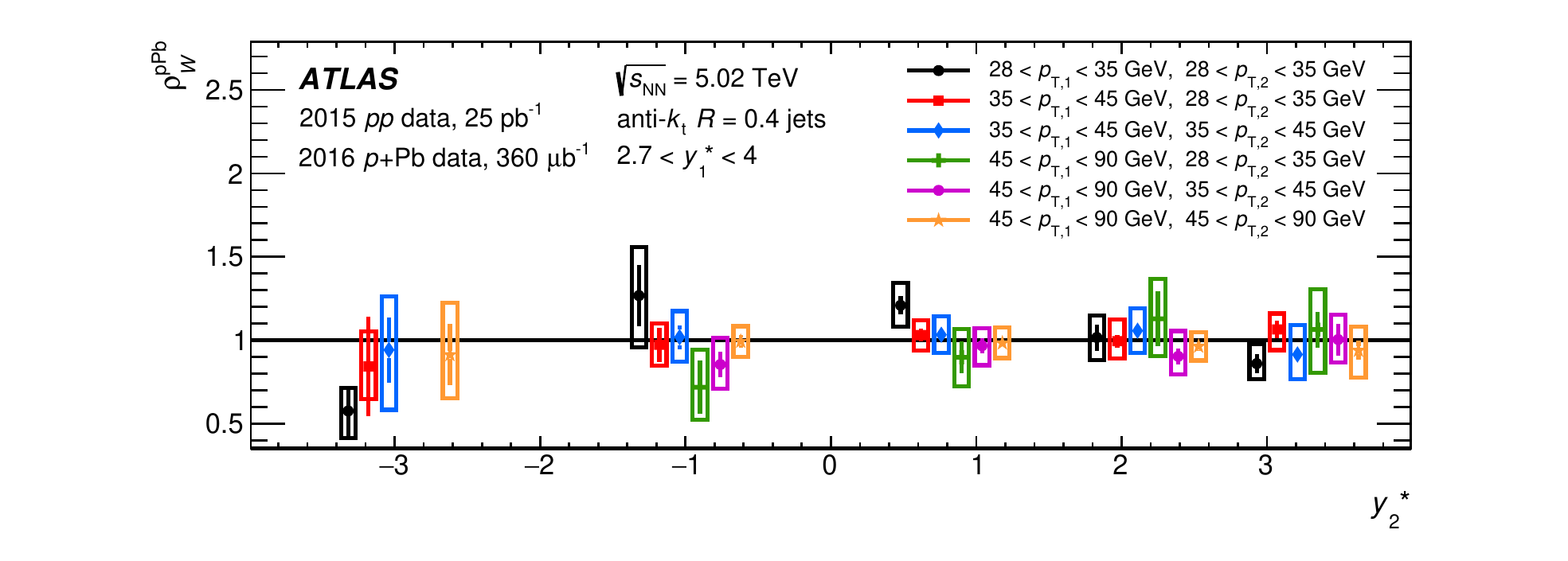} \\
			\includegraphics[width=1.2\textwidth]{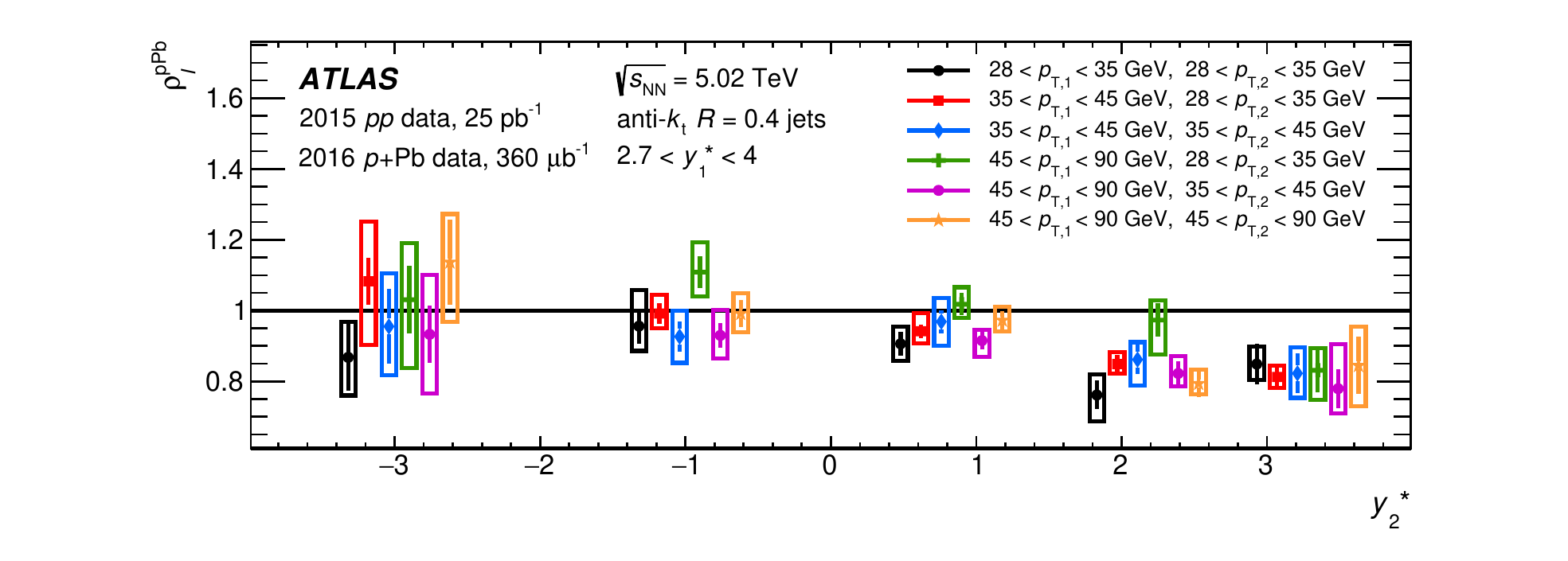} \\
		\end{tabular}
	}
	\caption{Ratios (top) \cppb\ of \wonetwo\ and (bottom) \ippb\ of \ionetwo\ values between \pPb\ collisions and \pp\ collisions for different selections of \ptone\ and \pttwo\ as a function of \ystartwo. The data points are shifted horizontally for visibility, and do not reflect an actual shift in rapidity. The vertical size of the open boxes represents systematic uncertainties and the error bars indicate statistical uncertainties. The horizontal size of the open boxes does not represent the width of the bins. Some points are not presented due to large statistical uncertainties. Results are shown with no $\Delta\pt$ requirement, where $\Delta\pt=\ptone-\pttwo$.}
	\label{fig:finalratios}
\end{figure}

The ratios \cppb\ between \pPb\ collisions and \pp\ collisions for different ranges of \ptone\ and \pttwo\ as a function of \ystartwo\ are  consistent with unity and are presented in the top panel of Figure~\ref{fig:finalratios}. The ratios \ippb\ between \pPb\ collisions and \pp\ collisions in the same bins of rapidity and transverse momentum are shown in the bottom panel of Figure~\ref{fig:finalratios}. The uncertainty of this ratio is dominated by systematic uncertainties, which are correlated in jet \pt\ and \ystar. The ratios \ippb\ are consistent with unity for subleading jets in the lead-going direction and for central--forward dijets. The ratio of conditional yields of jet pairs when both the leading and subleading jets are in the proton-going direction is suppressed by approximately 20\% in \pPb\ collisions compared to \pp\ collisions, with no significant dependence on jet \pt. In the most forward--forward configuration, with both jets in the lowest jet-\pt\ interval $28<\ptone, \pttwo< 35$~\GeV, the \xb\ range probed is between $10^{-4}$ and $10^{-3}$. The suppression indicates a reduction in the nuclear gluon density per nucleon relative to the unbound nucleon in a region where nuclear shadowing and saturation are predicted~\cite{vanHameren:2019ysa}.

\begin{figure}[ht]
	\centerline{
		\begin{tabular}{cc}
			\includegraphics[width=0.5\textwidth]{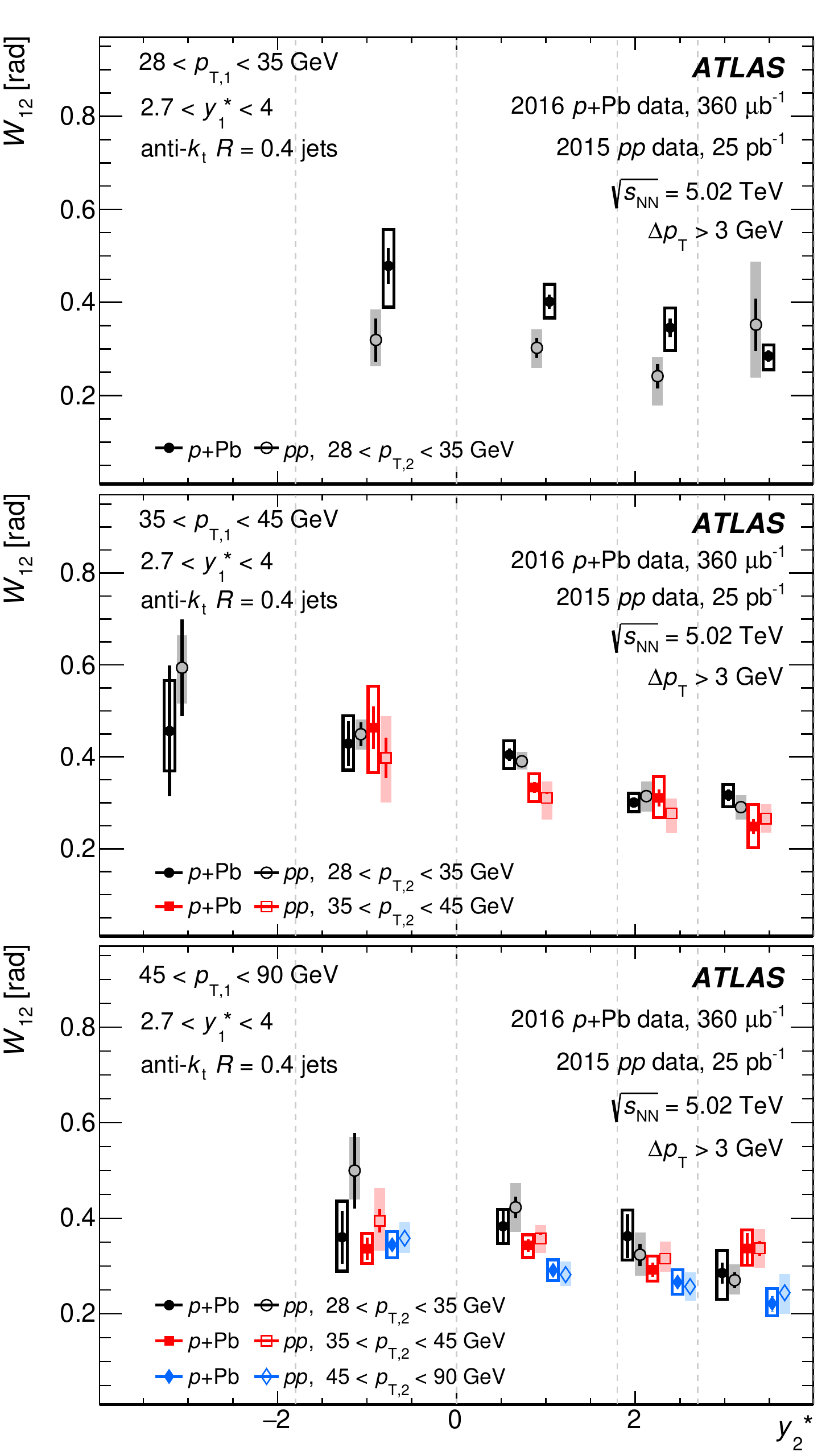} &
			\includegraphics[width=0.5\textwidth]{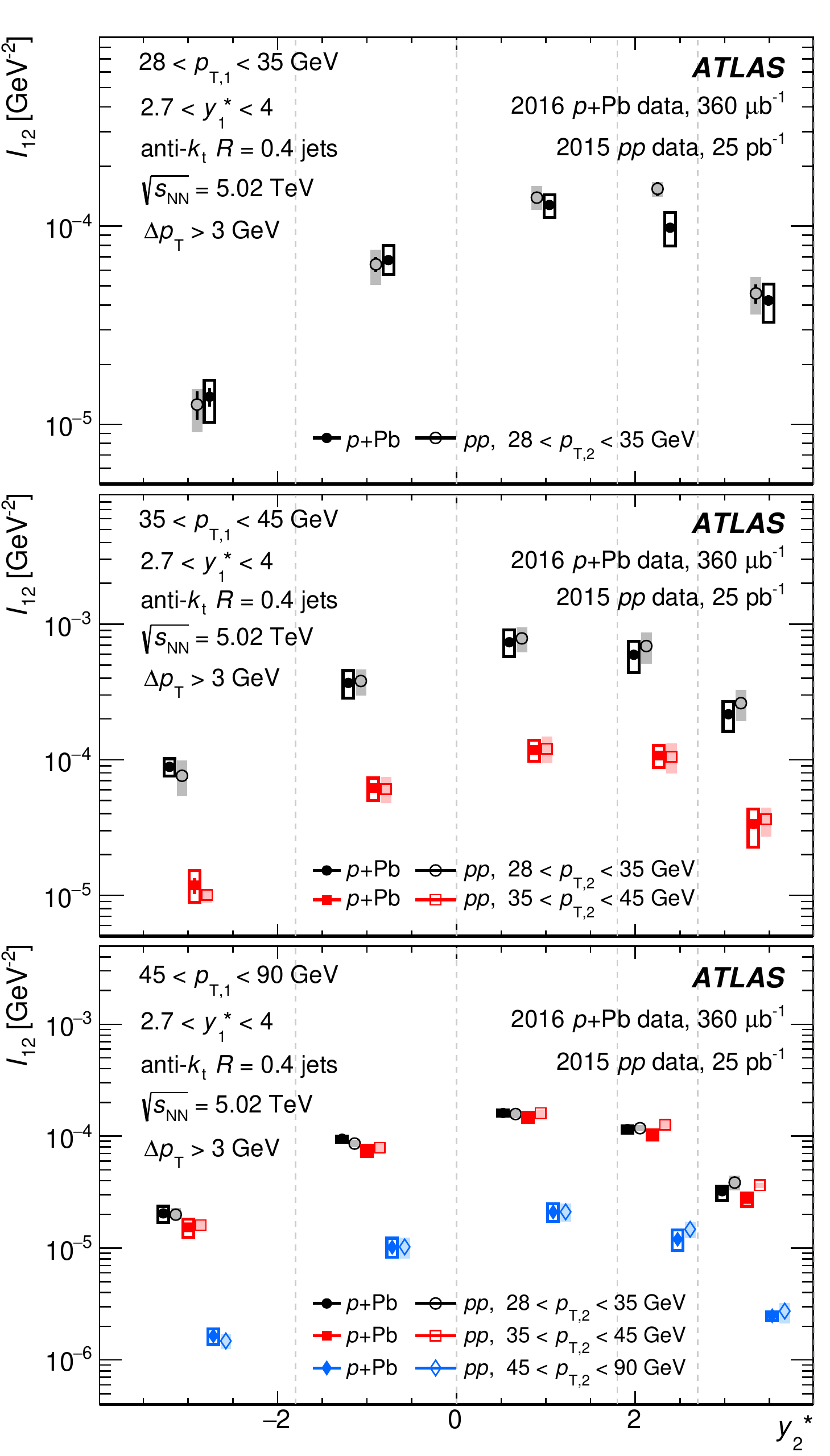} \\
		\end{tabular}
	}
	\caption{Comparison of (left) \wonetwo\ and (right) \ionetwo\ values in  \pp\ (open symbols) and \pPb\ (closed symbols) collisions for different selections of \ptone\ and \pttwo\ as a function of \ystartwo. The \ystartwo\ intervals are separated by dotted vertical lines. The data points are shifted horizontally for visibility, and do not reflect an actual shift in rapidity. The vertical size of the shaded and open boxes represents systematic uncertainties for \pp\ and \pPb, respectively, and the error bars indicate statistical uncertainties. The horizontal size of the shaded and open boxes does not represent the width of the bins. Some data points in the rapidity interval of  $-4.0 < \ystartwo < 1.8$ are not presented  due to large statistical uncertainties. Results are shown with the requirement of $\Delta\pt > 3$~\GeV, where $\Delta\pt=\ptone-\pttwo$.}
	\label{fig:finalplotswithcuts}
\end{figure}

\begin{figure}[ht]
	\centerline{
		\begin{tabular}{c}
			\includegraphics[width=1.2\textwidth]{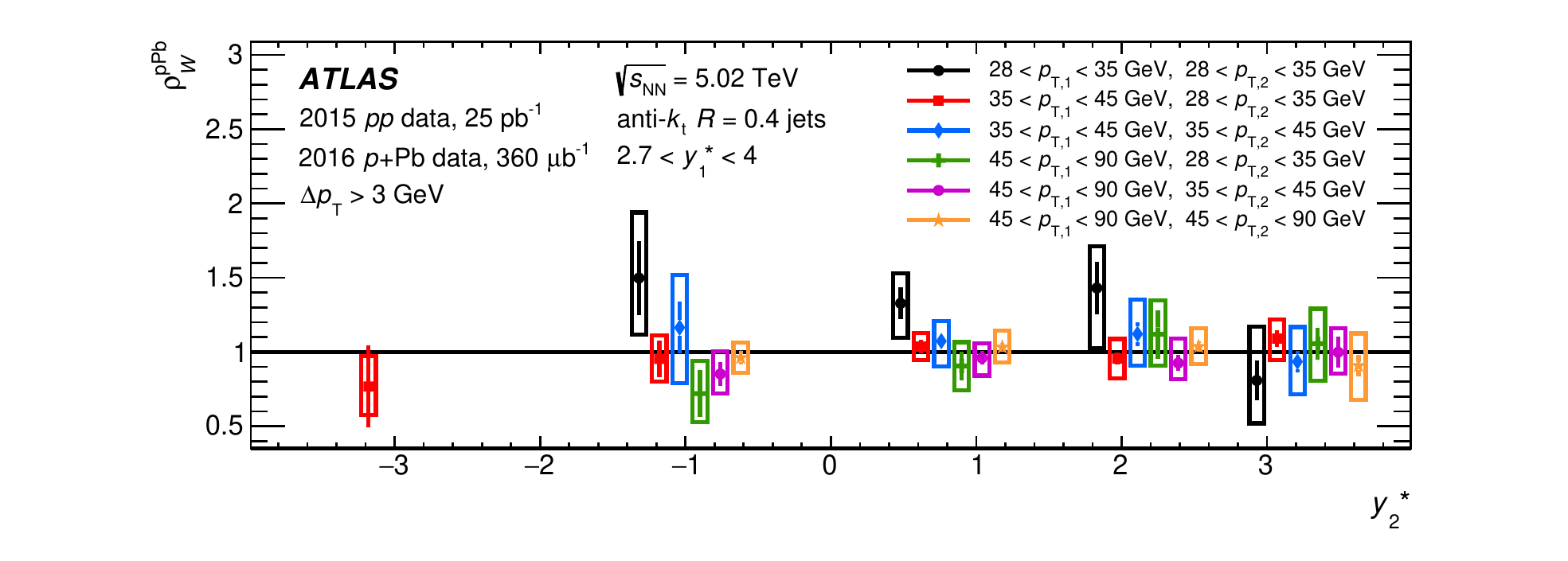} \\
			\includegraphics[width=1.2\textwidth]{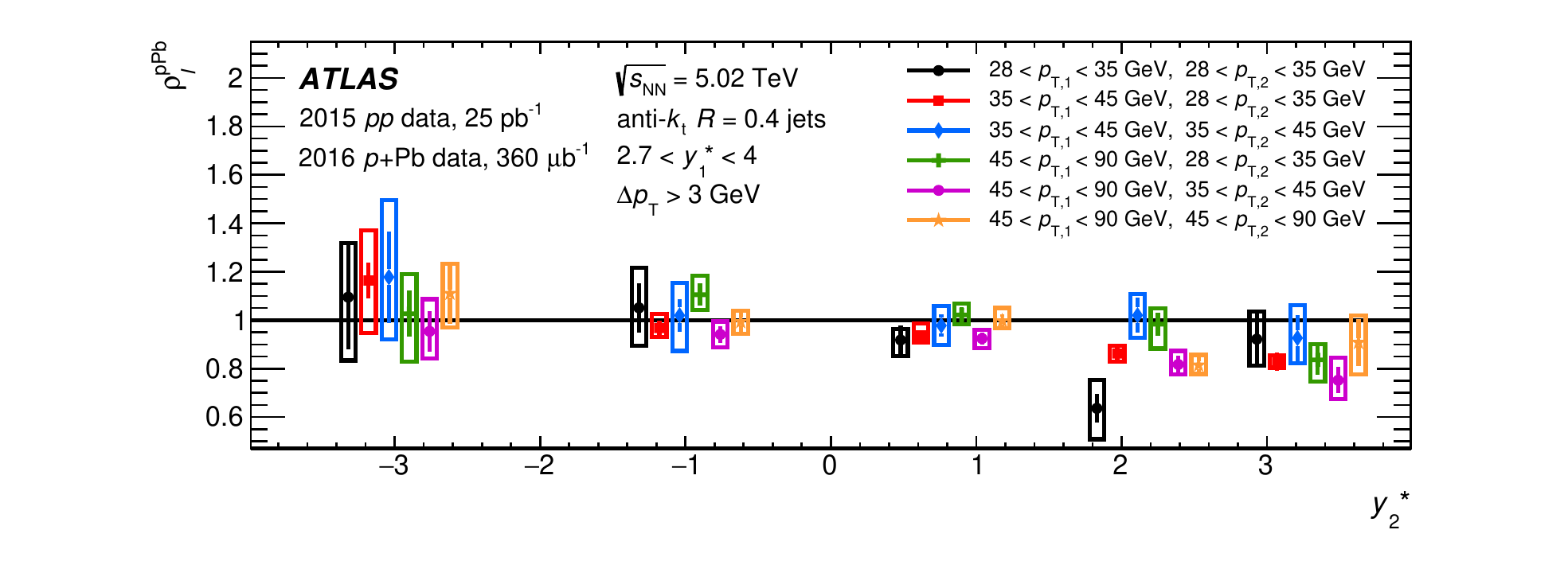} \\
		\end{tabular}
	}
	\caption{Ratios (top) \cppb\ of \wonetwo\  and (bottom) \ippb\ of \ionetwo\ values between \pPb\ collisions and \pp\ collisions for different selections of \ptone\ and \pttwo\ as a function of \ystartwo.  The data points are shifted horizontally for visibility, and do not reflect an actual shift in rapidity. The vertical size of the open boxes represents systematic uncertainties and the error bars indicate statistical uncertainties. The horiztonal size of the open boxes does not represent the width of the bin. Some data points in the rapidity interval of $-4.0 < \ystartwo < 1.8$ are not presented  due to large statistical uncertainties. Results are shown with the requirement of $\Delta\pt > 3$~\GeV, where $\Delta\pt=\ptone-\pttwo$.}
	\label{fig:finalratioswithcuts}
\end{figure}

Results for the values of \wonetwo\ and \ionetwo\ from \pp\ collisions and \pPb\ collisions with the requirement of $\Delta\pt>3$~\GeV\ are shown in Figure~\ref{fig:finalplotswithcuts}. The ratios of the two \wonetwo\ and \ionetwo\ values, \cppb\ and \ippb, are shown in Figure~\ref{fig:finalratioswithcuts}. The values of \wonetwo\ and \cppb\ are observed to be unaffected by the $\Delta\pt$  requirement. The conditional yields \ionetwo\ are smaller than the results with no $\Delta\pt$ requirement, while the conditional yield ratios \ippb\ are unaffected by the $\Delta\pt$ requirement.

\FloatBarrier

\section{Summary}
\label{sec:summary}
This paper presents measurements of dijet azimuthal angular correlations and the conditional yields of leading and subleading jets in \pp\ and \pPb\ collisions at $\sqrt{s_{_\text{NN}}}$~=~5.02~\TeV. The data, recorded by the ATLAS experiment at the Large Hadron Collider, correspond to 25~pb$^{-1}$ and 360~$\mu$b$^{-1}$ of $pp$ and \pPb\ collisions, respectively. The measurement utilizes pairs of \RFour\ \antikt\ jets in the transverse momentum range $28 < \pT < 90$~\GeV\ and center-of-mass rapidity range $-4.0 < y^{*} <4.0$. The shapes of the azimuthal angular correlation functions for forward--forward and forward--central dijets and conditional yields are sensitive to possible effects of gluon saturation at low \xb. Dijets with a large separation in rapidity and where both jets have small transverse momentum probe an approximate \xb\ range between $10^{-4}$ and $10^{-3}$.

The widths of the azimuthal correlation functions are found to be smaller for pairs of jets with higher $\ptone, \pttwo$, but larger for large rapidity interval between the jets. No significant broadening of azimuthal angular correlations is observed for forward--forward and forward--central dijets in \pPb\ compared to \pp\ collisions. The measurement of conditional yields of forward--forward dijets in \pPb\ collisions compared to \pp\ collisions shows a suppression of approximately 20\%, with no significant dependence on jet \pt. The observed suppression can be interpreted in terms of the nuclear gluon density in a low-$x$ region where it is not well known. It may therefore be used to constrain possible nuclear effects including saturation.

\clearpage

\section*{Acknowledgments}

We thank CERN for the very successful operation of the LHC, as well as the support staff from our institutions without whom ATLAS could not be operated efficiently.

We acknowledge the support of ANPCyT, Argentina; YerPhI, Armenia; ARC, Australia; BMWFW and FWF, Austria; ANAS, Azerbaijan; SSTC, Belarus; CNPq and FAPESP, Brazil; NSERC, NRC and CFI, Canada; CERN; CONICYT, Chile; CAS, MOST and NSFC, China; COLCIENCIAS, Colombia; MSMT CR, MPO CR and VSC CR, Czech Republic; DNRF and DNSRC, Denmark; IN2P3-CNRS, CEA-DRF/IRFU, France; SRNSFG, Georgia; BMBF, HGF, and MPG, Germany; GSRT, Greece; RGC, Hong Kong SAR, China; ISF and Benoziyo Center, Israel; INFN, Italy; MEXT and JSPS, Japan; CNRST, Morocco; NWO, Netherlands; RCN, Norway; MNiSW and NCN, Poland; FCT, Portugal; MNE/IFA, Romania; MES of Russia and NRC KI, Russian Federation; JINR; MESTD, Serbia; MSSR, Slovakia; ARRS and MIZ\v{S}, Slovenia; DST/NRF, South Africa; MINECO, Spain; SRC and Wallenberg Foundation, Sweden; SERI, SNSF and Cantons of Bern and Geneva, Switzerland; MOST, Taiwan; TAEK, Turkey; STFC, United Kingdom; DOE and NSF, United States of America. In addition, individual groups and members have received support from BCKDF, CANARIE, CRC and Compute Canada, Canada; COST, ERC, ERDF, Horizon 2020, and Marie Sk{\l}odowska-Curie Actions, European Union; Investissements d' Avenir Labex and Idex, ANR, France; DFG and AvH Foundation, Germany; Herakleitos, Thales and Aristeia programmes co-financed by EU-ESF and the Greek NSRF, Greece; BSF-NSF and GIF, Israel; CERCA Programme Generalitat de Catalunya, Spain; The Royal Society and Leverhulme Trust, United Kingdom. 

The crucial computing support from all WLCG partners is acknowledged gratefully, in particular from CERN, the ATLAS Tier-1 facilities at TRIUMF (Canada), NDGF (Denmark, Norway, Sweden), CC-IN2P3 (France), KIT/GridKA (Germany), INFN-CNAF (Italy), NL-T1 (Netherlands), PIC (Spain), ASGC (Taiwan), RAL (UK) and BNL (USA), the Tier-2 facilities worldwide and large non-WLCG resource providers. Major contributors of computing resources are listed in Ref.~\cite{ATL-GEN-PUB-2016-002}.

\printbibliography

\clearpage 

 
\begin{flushleft}
{\Large The ATLAS Collaboration}

\bigskip

M.~Aaboud$^\textrm{\scriptsize 35d}$,    
G.~Aad$^\textrm{\scriptsize 100}$,    
B.~Abbott$^\textrm{\scriptsize 127}$,    
D.C.~Abbott$^\textrm{\scriptsize 101}$,    
O.~Abdinov$^\textrm{\scriptsize 13,*}$,    
B.~Abeloos$^\textrm{\scriptsize 131}$,    
D.K.~Abhayasinghe$^\textrm{\scriptsize 92}$,    
S.H.~Abidi$^\textrm{\scriptsize 166}$,    
O.S.~AbouZeid$^\textrm{\scriptsize 40}$,    
N.L.~Abraham$^\textrm{\scriptsize 155}$
% ...
\bigskip
\\

% ...
$^{13}$Institute of Physics, Azerbaijan Academy of Sciences, Baku; Azerbaijan.\\
% ...
$^{35}$$^{(a)}$Facult\'e des Sciences Ain Chock, R\'eseau Universitaire de Physique des Hautes Energies - Universit\'e Hassan II, Casablanca;$^{(b)}$Facult\'{e} des Sciences, Universit\'{e} Ibn-Tofail, K\'{e}nitra;$^{(c)}$Facult\'e des Sciences Semlalia, Universit\'e Cadi Ayyad, LPHEA-Marrakech;$^{(d)}$Facult\'e des Sciences, Universit\'e Mohamed Premier and LPTPM, Oujda;$^{(e)}$Facult\'e des sciences, Universit\'e Mohammed V, Rabat; Morocco.\\
% ...
$^{40}$Niels Bohr Institute, University of Copenhagen, Copenhagen; Denmark.\\
% ...
$^{92}$Department of Physics, Royal Holloway University of London, Egham; United Kingdom.\\
% ...
$^{100}$CPPM, Aix-Marseille Universit\'e, CNRS/IN2P3, Marseille; France.\\
$^{101}$Department of Physics, University of Massachusetts, Amherst MA; United States of America.\\
% ...
$^{127}$Homer L. Dodge Department of Physics and Astronomy, University of Oklahoma, Norman OK; United States of America.\\
% ...
$^{131}$LAL, Universit\'e Paris-Sud, CNRS/IN2P3, Universit\'e Paris-Saclay, Orsay; France.\\
% ...
$^{155}$Department of Physics and Astronomy, University of Sussex, Brighton; United Kingdom.\\
% ...
$^{166}$Department of Physics, University of Toronto, Toronto ON; Canada.\\
% ...
$^{a}$ Also at Borough of Manhattan Community College, City University of New York, New York NY; United States of America.\\
$^{b}$ Also at Centre for High Performance Computing, CSIR Campus, Rosebank, Cape Town; South Africa.\\
$^{c}$ Also at CERN, Geneva; Switzerland.\\
$^{d}$ Also at CPPM, Aix-Marseille Universit\'e, CNRS/IN2P3, Marseille; France.\\
$^{e}$ Also at D\'epartement de Physique Nucl\'eaire et Corpusculaire, Universit\'e de Gen\`eve, Gen\`eve; Switzerland.\\
% ...
$^{*}$ Deceased

\end{flushleft}


\end{document}